\documentclass[twocolumn]{aastex631}

\newcommand\tess{TESS}
\newcommand\gaia{\textit{Gaia}}
\newcommand\kms{$\textrm{km~s}^{-1}$}
\newcommand\ms{$\textrm{m~s}^{-1}$}

\newcommand\gcmcubed{$\textrm{g~cm}^{-3}$}
\usepackage[T1]{fontenc} 
\usepackage[utf8]{inputenc} 
\usepackage[stable]{footmisc}
\newcommand\xx{\textbf{FINDME}}

\newcommand{\unit}[1]{\ensuremath{\, \mathrm{#1}}} 

\newcommand\solmass{$M_{\odot}$}
\newcommand\solradius{$R_{\odot}$}
\newcommand{\UCI}{Department of Physics \& Astronomy, The University of California, Irvine, Irvine, CA 92697, USA}

\submitjournal{AAS Journals}

\begin{document}

\title{TOI-4201: An Early M-dwarf Hosting a Massive Transiting Jupiter Stretching Theories of Core-Accretion\footnote{This paper includes data gathered with the 6.5 meter Magellan Telescopes located at Las Campanas Observatory, Chile.}}

\shortauthors{Delamer et al. 2023}
\shorttitle{TOI-4201: M-dwarf with Massive Jupiter}

\author[0000-0001-8401-4300]{Megan Delamer}
\affil{Department of Astronomy \& Astrophysics, 525 Davey Laboratory, The Pennsylvania State University, University Park, PA 16802, USA}
\affil{Center for Exoplanets and Habitable Worlds, 525 Davey Laboratory, The Pennsylvania State University, University Park, PA 16802, USA}

\author[0000-0001-8401-4300]{Shubham Kanodia}
\affil{Earth and Planets Laboratory, Carnegie Institution for Science, 5241 Broad Branch Road, NW, Washington, DC 20015, USA}

\author[0000-0003-4835-0619]{Caleb I. Ca\~nas}
\altaffiliation{NASA Postdoctoral Fellow}
\affiliation{NASA Goddard Space Flight Center, 8800 Greenbelt Road, Greenbelt, MD 20771, USA }

\author[0000-0002-8278-8377]{Simon M\"{u}ller}
\affil{Center for Theoretical Astrophysics \& Cosmology, University of Zurich, Winterthurerstr. 190, CH-8057 Zurich, Switzerland}
\author[0000-0001-5555-2652]{Ravit Helled}
\affil{Center for Theoretical Astrophysics \& Cosmology, University of Zurich, Winterthurerstr. 190, CH-8057 Zurich, Switzerland}

\author[0000-0002-9082-6337]{Andrea S.J.\ Lin}
\affil{Department of Astronomy \& Astrophysics, 525 Davey Laboratory, The Pennsylvania State University, University Park, PA, 16802, USA}
\affil{Center for Exoplanets and Habitable Worlds, 525 Davey Laboratory, The Pennsylvania State University, University Park, PA, 16802, USA}
\author[0000-0002-2990-7613]{Jessica E.\ Libby-Roberts}
\affil{Department of Astronomy \& Astrophysics, 525 Davey Laboratory, The Pennsylvania State University, University Park, PA, 16802, USA}
\affil{Center for Exoplanets and Habitable Worlds, 525 Davey Laboratory, The Pennsylvania State University, University Park, PA, 16802, USA}
\author[0000-0002-5463-9980]{Arvind F.\ Gupta}
\affil{Department of Astronomy \& Astrophysics, 525 Davey Laboratory, The Pennsylvania State University, University Park, PA 16802, USA}
\affil{Center for Exoplanets and Habitable Worlds, 525 Davey Laboratory, The Pennsylvania State University, University Park, PA 16802, USA}
\author[0000-0001-9596-7983]{Suvrath Mahadevan}
\affil{Department of Astronomy \& Astrophysics, 525 Davey Laboratory, The Pennsylvania State University, University Park, PA 16802, USA}
\affil{Center for Exoplanets and Habitable Worlds, 525 Davey Laboratory, The Pennsylvania State University, University Park, PA 16802, USA}

\author[0009-0008-2801-5040]{Johanna Teske}
\affil{Earth and Planets Laboratory, Carnegie Institution for Science, 5241 Broad Branch Road, NW, Washington, DC 20015, USA}
\author[0000-0003-1305-3761]{R.\ Paul Butler}
\affil{Earth and Planets Laboratory, Carnegie Institution for Science, 5241 Broad Branch Road, NW, Washington, DC 20015, USA}
\author[0000-0001-7961-3907]{Samuel W.\ Yee}
\affiliation{Department of Astrophysical Sciences, Princeton University, 4 Ivy Lane, Princeton, NJ 08544, USA}

\author[0000-0002-5226-787X]{Jeffrey D. Crane}
\affil{The Observatories of the Carnegie Institution for Science, 813 Santa Barbara Street, Pasadena, CA, 91101}
\author[0000-0002-8681-6136]{Stephen Shectman}
\affil{The Observatories of the Carnegie Institution for Science, 813 Santa Barbara Street, Pasadena, CA, 91101}
\author[0000-0003-0412-9664]{David Osip}
\affil{Las Campanas Observatory, Carnegie Institution for Science, Colina el Pino, Casilla 601 La Serena, Chile}
\author{Yuri Beletsky}
\affil{Las Campanas Observatory, Carnegie Institution for Science, Colina el Pino, Casilla 601 La Serena, Chile}

\author[0000-0002-0048-2586]{Andrew Monson}
\affil{Steward Observatory, The University of Arizona, 933 N. Cherry Ave, Tucson, AZ 85721, USA}
\author[0000-0003-0353-9741]{Jaime A. Alvarado-Montes}
\affil{School of Mathematical and Physical Sciences, Macquarie University, Balaclava Road, North Ryde, NSW 2109, Australia}
\affil{The Macquarie University Astrophysics and Space Technologies Research Centre, Macquarie University, Balaclava Road, North Ryde, NSW 2109, Australia}
\author[0000-0003-4384-7220]{Chad F.\ Bender}
\affil{Steward Observatory, The University of Arizona, 933 N. Cherry Ave, Tucson, AZ 85721, USA}
\author[0000-0002-3610-6953]{Jiayin Dong} 
\altaffiliation{Flatiron Research Fellow} 
\affil{Center for Computational Astrophysics, Flatiron Institute, New York, NY, USA}

\author[0000-0002-7127-7643]{Te Han}
\affil{\UCI}


\author[0000-0001-8720-5612]{Joe P.\ Ninan}
\affil{Department of Astronomy and Astrophysics, Tata Institute of Fundamental Research, Homi Bhabha Road, Colaba, Mumbai 400005, India}

\author[0000-0003-0149-9678]{Paul Robertson}
\affil{\UCI}

\author[0000-0001-8127-5775]{Arpita Roy}
\affil{Space Telescope Science Institute, 3700 San Martin Drive, Baltimore, MD 21218, USA}
\affil{Department of Physics and Astronomy, Johns Hopkins University, 3400 N Charles St, Baltimore, MD 21218, USA}

\author[0000-0002-4046-987X]{Christian Schwab}
\affil{School of Mathematical and Physical Sciences, Macquarie University, Balaclava Road, North Ryde, NSW 2109, Australia}

\author[0000-0001-7409-5688]{Guðmundur Stefánsson} 
\affil{NASA Sagan Fellow}
\affil{Department of Astrophysical Sciences, Princeton University, 4 Ivy Lane, Princeton, NJ 08540, USA}
\author[0000-0001-6160-5888]{Jason T.\ Wright}
\affil{Department of Astronomy \& Astrophysics, 525 Davey Laboratory, The Pennsylvania State University, University Park, PA, 16802, USA}
\affil{Center for Exoplanets and Habitable Worlds, 525 Davey Laboratory, The Pennsylvania State University, University Park, PA, 16802, USA}
\affil{Penn State Extraterrestrial Intelligence Center, 525 Davey Laboratory, The Pennsylvania State University, University Park, PA, 16802, USA}

\correspondingauthor{Megan Delamer}
\email{mmd6393@psu.edu}

\begin{abstract}

We confirm TOI-4201~b as a transiting Jovian mass planet orbiting an early M dwarf discovered by the Transiting Exoplanet Survey Satellite. Using ground based photometry and precise radial velocities from NEID and the Planet Finder Spectrograph, we measure a planet mass of 2.59$^{+0.07}_{-0.06}$\,M$_{J}$, making this one of the most massive planets transiting an M-dwarf. The planet is $\sim$0.4\% the mass of its 0.63~\solmass~host and may have a heavy element mass comparable to the total dust mass contained in a typical Class II disk. TOI-4201~b stretches our understanding of core-accretion during the protoplanetary phase, and the disk mass budget, necessitating giant planet formation to either take place much earlier in the disk lifetime, or perhaps through alternative mechanisms like gravitational instability.

\end{abstract}


\section{Introduction} \label{sec:intro}

The Transiting Exoplanet Survey Satellite \citep[TESS;][]{ricker_transiting_2014} observes millions of stars across the entire sky searching for transiting candidates. It has been instrumental in finding planets around M-dwarf stars, the most common type of star in the Galaxy.  Among the confirmed TESS candidate planets orbiting M-dwarfs are $\sim$15 giant planets in short period orbits \citep[e.g.,][]{gan_toi-530b_2022,kanodia_toi-5205b_2023, hobson_toi-3235_2023, kagetani_mass_2023}, which seem to challenge current theories of planet formation through core accretion \citep{laughlin_core_2004, ida_toward_2005}.

Under the core-accretion model, the formation of giant exoplanets around M-dwarf stars (GEMS) is difficult for two main reasons. First, we expect disk mass, particularly the solid surface density, to scale with host star mass \citep{andrews_mass_2013, pascucci_steeper_2016}. Under typical opacity assumptions runaway gas accretion can only be triggered after a core of $\sim$\,10\,M$_{\oplus}$ is formed.  Since 10\,M$_{\oplus}$ is a large percentage of the solid mass typically available in a disk around an M-dwarf \citep{ansdell_alma_2016, Tazzari_3mm_Lup_survey_2021, manara_demographics_2022}, forming this core would be a challenge.  Second, due to their lower host star masses, the Keplerian orbital time-scales are much longer for M-dwarfs compared to solar-type stars, which coupled with their lower disk surface densities should make it harder to form a massive enough core in time to initiate runaway gaseous accretion before the disk dissipates \citep{laughlin_core_2004}.

The discovery and characterization of these GEMS is already challenging our understanding of giant planet formation, and despite their enhanced detection probabilities with transit and radial velocity observations (large planet to stellar mass and radius ratios), they remain rare. While currently difficult to perform sample level comparisons on the existing population ($<$ 15), efforts are underway to significantly increase this sample size and to enable such analysis in the future.

In this letter, we present the confirmation of a massive super-Jupiter around the early M-dwarf TOI-4201. While previous gas giants have been difficult to explain from a mass budget perspective \citep{kanodia_toi-5205b_2023, hobson_toi-3235_2023}, one proposed explanation is that planet formation begins early on in disk lifetime ($<$ 1 Myr), when the disks are much more massive. The discovery of a super-Jupiter around an early M-dwarf with a planet to host mass ratio of $\sim$0.4\% stretches this even further.

To characterize TOI-4201 and confirm the TESS signal as a planet, we use precision radial velocities (RVs) from the Planet Finder Spectrograph (PFS) and NEID, high contrast speckle imaging from WIYN/NESSI, and ground based photometry from TMMT and LCRO, as well as archival photometry from LCOGT.  We detail these observations in Section \ref{sec:observations} and describe the stellar characterization in Section \ref{sec:stellar}.  Section \ref{sec:joint} covers the data analysis, including details of the joint modeling of RVs and photometry.  In Section \ref{sec:discussion}, we compare TOI-4201~b with other giant planets around M-dwarfs and consider potential formation scenarios. We summarize our findings in Section \ref{sec:conclusion}.

\section{Observations}\label{sec:observations}
\subsection{TESS}\label{sec:TESS}

TOI-4201 (TIC 95057860, Gaia DR3 2997312063605005056) was identified as hosting a transiting object of interest in the \tess~ Sector 6 long-cadence (1800 s) light curve spanning 2018 December 11 to 2019 January 7, by the Quick Look Pipeline \citep[QLP;][]{huang_photometry_2020} during the \textit{Faint-Star Search} \citep{kunimoto_tess_2022}. It was re-observed by \tess~ during Sector 33 from 2020 December 17 to 2021 January 13 with 600 s candence.  We extract a light curve for each sector by performing aperture photometry with \texttt{eleanor} \citep{feinstein_eleanor_2019} using the \texttt{CORR\_FLUX} values, in which \texttt{eleanor} uses linear regression with pixel position, measured background, and time to remove signals correlated with these parameters. We show the \tess~ photometry in \autoref{fig:orbitfig}.

\subsection{Ground based photometric follow up}\label{sec:photometry}

\subsubsection{TMMT \& LCRO}

We observed a transit of TOI-4201~b on the night of 2021 December 28 with the Three-hundred MilliMeter Telescope \citep[TMMT;][]{monson_standard_2017} at Las Campanas Observatory, with 120~s exposures in Bessell I. Observations were taken as the target set from an airmass of 1.19 to 2.48.

We then observed a second transit on the night of 2022 January 15, with TMMT in Bessell V and the 0.6~m Las Campanas Remote Observatory (LCRO) in SDSS $i^\prime$, both using 300~s exposures. The target set from an airmass of 1.04 to 1.66 over the course of the night. 

We reduced the data from all three observations (bias correction, flat-fielding, cosmic/bad-pixel removal, etc.) following the procedure outlined in \citet{monson_standard_2017}. Then, we performed differential aperture photometry on all images using a python script based on \citet{monson_standard_2017}. Final light curves are plotted in \autoref{fig:orbitfig}.


\subsubsection{1.0 m LCOGT Archival Data}

We also retrieve a full transit for TOI-4201~b from the Las Cumbres Observatory global telescope network \citep[LCOGT;][]{brown_2013_LCOGT} public data archive. This transit was observed in both SDSS $g^\prime$ and $i^\prime$ on the night of 2021 October 3 (proposal ID: KEY2020B-005, PI: Shporer, A.) from the Sinistro imaging camera on the 1m LCOGT Dome B telescope at Cerro Tololo Inter-American Observatory. These observations were taken mildly defocused (FWHM $\sim$2.5\arcsec) at exposure times of 300~s in $g^\prime$ and 180~s in $i^\prime$, with the target rising from an airmass of $\sim$2.3 to 1.06 over the course of the night. The raw data were automatically processed by the BANZAI pipeline \citep{mccully_2018_BANZAI}. We then perform aperture photometry on the processed images using \texttt{AstroImageJ} \citep[\texttt{AIJ};][]{collins_astroimagej_2017}. We see a strong slope in both the $g^\prime$ and $i^\prime$ lightcurves, so we detrend both datasets, in time, prior to our analysis. These transits are shown in Panel e and f of \autoref{fig:orbitfig}.

\subsection{High Contrast Imaging}

\subsubsection{NESSI}

We observed TOI-4201 on the night of 2022 April 18 with the NN-Explore Exoplanet Stellar Speckle Imager \citep[NESSI;][]{scott_nn-explore_2018} on the WIYN\footnote{The WIYN Observatory is a joint facility of the NSF's National Optical-Infrared Astronomy Research Laboratory, Indiana University, the University of Wisconsin-Madison, Pennsylvania State University, the University of Missouri, the University of California-Irvine, and Purdue University.} 3.5\,m telescope at Kitt Peak National Observatory to rule out stellar companions. We took diffraction-limited exposures using the red camera and the Sloan \(z^\prime\) filter at a 40\,ms cadence for 9\,minutes and reconstructed the speckle image following the methods described by \citet{howell_speckle_2011}. We compute $5\sigma$ contrast limits as a function of separation, $\Delta\theta$, from the primary source and find no evidence for nearby or background sources with $\Delta$\(z^\prime\) $<3.5$ at $\Delta\theta=0.5$'' and $\Delta$\(z^\prime\) $<3.9$ at $\Delta\theta=1.2$''.

\subsection{Radial velocity follow-up}\label{sec:neidrvs}

\begin{deluxetable}{cccc}
\tablecaption{RVs for TOI-4201. \label{tab:rvs}}
\tablehead{\colhead{$\unit{BJD_{TDB}}$}  &  \colhead{RV}   & \colhead{$\sigma$}  & \colhead{Instrument} \\
           \colhead{(d)}   &  \colhead{\ms{}} & \colhead{\ms{}} & \colhead{}}
\startdata
2459898.89219 & 124.5 & 61.1 & NEID \\
2459899.87168 & -82.3 & 49.4 & NEID \\
2459903.87132 & -450.0 & 67.3 & NEID \\
2459983.56562 & -390.7 & 13.9 & PFS \\ 
2459983.57882 & -400.1 & 11.5 & PFS \\ 
2459983.59321 & -386.2 & 14.1 & PFS \\ 
2459984.55132 & 263.3 & 15.5 & PFS \\ 
2459984.56623 & 259.4 & 16.6 & PFS \\ 
2459984.58013 & 304.9 & 16.3 & PFS \\ 
2460042.51200 & 509.6 & 10.2 & PFS \\ 
2460042.52631 & 499.5 & 12.8 & PFS \\ 
2460043.50692 & 0.0 & 9.7 & PFS \\ 
2460043.52043 & -32.1 & 8.6 & PFS \\ 
2460043.53418 & -9.1 & 8.2 & PFS \\ 
2460044.51157 & -431.8 & 8.9 & PFS \\ 
2460044.52556 & -410.4 & 9.5 & PFS \\ 
2460044.53970 & -413.8 & 10.6 & PFS \\ 
2460046.51302 & 371.3 & 10.0 & PFS \\ 
2460046.52716 & 374.8 & 9.8 & PFS \\ 
2460046.54106 & 358.9 & 9.6 & PFS \\ 
\enddata
\end{deluxetable}
\subsubsection{NEID}

We observed TOI-4201 using NEID, a fiber-fed, environmentally stabilized  spectrograph on the WIYN 3.5 m telescope \citep{halverson_comprehensive_2016, schwab_design_2016, robertson_ultrastable_2019} for three epochs between 2022 November 15 and 2022 November 20. NEID covers the wavelengths 380 to 930 nm with a spectral resolution of $R \sim 110,000$. Each visit consisted of a single 30 minute exposure. We use the standard NEID data reduction pipeline followed by the \texttt{SERVAL} algorithm developed by \cite{zechmeister_spectrum_2018}, and adapted for NEID by \cite{stefansson_warm_2022}. At 850\,nm, the SNR for each of the three RV points is 6.0, 7.6, and 5.5; we attribute the high jitter seen in the fit to the low SNR. The final NEID RVs are included in \autoref{tab:rvs} and as a machine readable table with the manuscript.

\subsubsection{Planet Finder Spectrograph}
We observed TOI-4201 with the Planet Finder Spectrograph \citep[PFS;][]{crane_carnegie_2006, crane_carnegie_2008, crane_carnegie_2010} on the 6.5 m Magellan II (Clay) telescope at Las Campanas Observatory. Between 2022 November 8 and 2023 April 12 we obtained 6 visits, each consisting of three exposures of 1200 seconds\footnote{Barring the visit on 2023 April 8 (BJD 2460042.512), which had to be cut short after two exposures due to adverse weather.} in 3 x 3 CCD binning mode with a 0.3\arcsec slit. This data was taken with the iodine gas absorption cell in the light path, which imprints a dense forest of molecular lines \citep{hatzes_iodine_2019} between 5000 to 6200 \AA. We also obtained one template spectrum without the iodine cell consisting of six exposures of 1200 seconds. The RVs were derived following the methodology of \cite{butler_attaining_1996}. As noted by \cite{hartman_hats-6b_2015} and \cite{bakos_hats-71b_2020}, due to the faintness of the target, and the optical region for the iodine region, the formal errors on the PFS RVs are likely underestimated. The final PFS RVs are included in \autoref{tab:rvs} and as a machine readable table with the manuscript.

\section{Stellar Parameters}\label{sec:stellar}%
The stellar parameters presented in \autoref{tab:stellarparam} are derived using the available broadband photometry and \textit{Gaia} astrometry. The stellar metallicity is estimated as $\mathrm{[Fe/H]}=0.30\pm0.15$ using WISE and Gaia colors\footnote{See \url{https://chrduque.shinyapps.io/metamorphosis/}} \citep[Equation 4 from][]{duque-arribas_photometric_2023}. This photometric relationship was determined by \cite{duque-arribas_photometric_2023} to be the most accurate photometric calibration when compared to a well-characterized  (i) spectroscopic sample of M-dwarfs \citep{birky_temperatures_2020} and (ii) M-dwarfs in binary systems \citep{Montes2018}. The stellar radius is calculated as $R_\star=0.62\pm0.02$~\solradius~using Equation 5 from \cite{mann_how_2015}. This value was then used to determine a mass of $M_\star=0.63\pm0.02$ \solmass~with Equation 6 from \cite{schweitzer_carmenes_2019}, which also agrees with the stellar mass estimate from \cite{mann_how_2019}. The stellar effective temperature is derived as $3920 \pm 50$ K with the empirical calibration from Equation 7 in \cite{rabus_discontinuity_2019}, that was derived using interferometric observations of well-characterized M-dwarfs. The adopted stellar parameters are consistent at the $1\sigma$ level with the (i) Bayesian stellar parameters ($\mathrm{[Fe/H]}=0.300_{-0.388}^{+0.002}$, $M_\star=0.65_{-0.050}^{+0.002}~\mathrm{M_\odot}$, $T_{\mathrm{eff}}=3955_{-49}^{+20}$ K) derived using the \texttt{StarHorse} code and a combination of multi-wavelength photometry and Gaia parallaxes \citep{anders_photo-astrometric_2022} and (ii) stellar parameters derived from a fit to the spectral energy distribution using \texttt{EXOFASTv2} \citep[using the broadband photometry in \autoref{tab:stellarparam};][]{eastman_exofastv2_2019}  ($M_\star=0.62\pm0.03~\mathrm{M_\odot}$, $R_\star=0.60\pm0.02~\mathrm{R_\odot}$, $T_{\mathrm{eff}}=3890\pm70$ K).

\subsection{Galactic kinematics}
We calculate the \textit{UVW} velocities in the barycentric frame and relative to the local standard of rest from \cite{schonrich_local_2010} using \texttt{GALPY} \citep{bovy_galpy_2015}\footnote{Following the convention of \textit{U} towards the Galactic center, \textit{V} towards the direction of Galactic spin, and \textit{W} towards the North Galactic Pole}. These velocities are reported in \autoref{tab:stellarparam} and are used to classify TOI-4201 as a thin disk star using the criteria from \cite{bensby_exploring_2014}. TOI-4201 is also determined to be a field star ($>99\%$ change of membership) using the BANYAN~$\Sigma$ tool \citep{gagne_banyan_2018}.

\begin{deluxetable*}{lccc}
\tablecaption{Summary of stellar parameters for TOI-4201. \label{tab:stellarparam}}
\tablehead{\colhead{~~~Parameter}&  \colhead{Description}&
\colhead{Value}&
\colhead{Reference}}
\startdata
\multicolumn{4}{l}{\hspace{-0.2cm} Main identifiers:}  \\
~~~TOI & \tess{} Object of Interest & 4201 & \tess{} mission \\
~~~TIC & \tess{} Input Catalogue  & 95057860 & Stassun \\
~~~2MASS & \(\cdots\) & J06015391-1327410 & 2MASS  \\
~~~Gaia DR3 & \(\cdots\) & 2997312063605005056 & Gaia DR3\\
\multicolumn{4}{l}{\hspace{-0.2cm} Equatorial Coordinates and Proper Motion:} \\
~~~$\alpha_{\mathrm{J2000}}$ &  Right Ascension (RA) & 90.475$\pm0.014$ & Gaia DR3\\
~~~$\delta_{\mathrm{J2000}}$ &  Declination (Dec) & -13.461$\pm0.015$ & Gaia DR3\\
~~~$\mu_{\alpha}$ &  Proper motion (RA, \unit{mas/yr}) &  11.731$\pm0.017$ & Gaia DR3\\
~~~$\mu_{\delta}$ &  Proper motion (Dec, \unit{mas/yr}) & 6.053$\pm0.018$ & Gaia DR3 \\
~~~$\varpi$ & Parallax & 5.291$\pm0.019$ & Gaia DR3\\
~~~$d$ &  Distance in pc  & $187.5_{-0.7}^{+0.6}$ & Bailer-Jones \\
\multicolumn{4}{l}{\hspace{-0.2cm} Optical and near-infrared magnitudes:}  \\
~~~$B$ & Johnson B mag & $16.69 \pm 0.15$ & APASS\\
~~~$V$ & Johnson V mag & $15.28 \pm 0.04$ & APASS\\
~~~$g^{\prime}$ &  Sloan $g^{\prime}$ mag  & $16.00 \pm 0.05$ & APASS\\
~~~$r^{\prime}$ &  Sloan $r^{\prime}$ mag  & $14.67 \pm 0.08$ & APASS \\
~~~$i^{\prime}$ &  Sloan $i^{\prime}$ mag  & $13.91 \pm 0.11$ & APASS \\
~~~$J$ & $J$ mag & $12.258 \pm 0.021$ & 2MASS\\
~~~$H$ & $H$ mag & $11.564 \pm 0.024$ & 2MASS\\
~~~$K_s$ & $K_s$ mag & $11.368 \pm 0.025$ & 2MASS\\
~~~$W1$ &  WISE1 mag & $11.272 \pm 0.024$ & WISE\\
~~~$W2$ &  WISE2 mag & $11.301 \pm 0.021$ & WISE\\
~~~$W3$ &  WISE3 mag & $11.283 \pm 0.155$ & WISE\\
\multicolumn{4}{l}{\hspace{-0.2cm} Photometric Relations:}\\
~~~$T_{\mathrm{eff}}$ &  Effective temperature in \unit{K} & $3920\pm50$ & This work\\
~~~$\mathrm{[Fe/H]}$ & Metallicity in dex & $0.30\pm0.15$ & This work \\
~~~$M_\star$ &  Mass in $M_{\odot}$ & $0.63\pm0.02$ & This work\\
~~~$R_\star$ &  Radius in $R_{\odot}$ & $0.62\pm0.02$ & This work\\
\multicolumn{4}{l}{\hspace{-0.2cm} Other Stellar Parameters:}           \\
~~~$\log(g)$ &  Surface gravity in cgs units & $4.65\pm0.03$ & This work \\
~~~$L_\star$ &  Luminosity in $L_{\odot}$ & $0.081\pm0.007$ & This work\\
~~~$\rho_\star$ &  Density in $\unit{g/cm^{3}}$ & $3.7\pm0.4$ & This work\\
~~~$\Delta$RV &  ``Absolute'' radial velocity in \unit{km/s} & $42.07 \pm 0.14$ & This work\\
~~~$U, V, W$ &  Galactic velocities in \unit{km/s} &  $-34.0 \pm 0.1, -27.46 \pm 0.09, -0.97 \pm 0.06$ & This work\\
~~~$U, V, W^a$ &  Galactic velocities (LSR) in \kms{} & $-22.9 \pm 0.8, -15.2 \pm 0.5, 6.3 \pm 0.4$ & This work\\
\enddata
\tablenotetext{}{References are: Stassun \citep{stassun_tess_2018}, 2MASS \citep{cutri_2mass_2003}, Gaia DR3 \citep{gaia_collaboration_gaia_2022}, Bailer-Jones \citep{bailer-jones_estimating_2021}, APASS \citep{henden_apass_2018}, WISE \citep{wright_wide-field_2010}}
\tablenotetext{a}{The barycentric UVW velocities are converted into local standard of rest (LSR) velocities using the constants from \cite{schonrich_local_2010}.}

\end{deluxetable*}

\section{Joint Fitting of Photometry and RVs}\label{sec:joint}

We perform a joint fit of the radial velocity time series and photometry using \texttt{exoplanet} \citep{foreman-mackey_exoplanet-devexoplanet_2021}, which utilizes \texttt{PyMC3} to perform Hamiltonian Monte Carlo (HMC) posterior sampling algorithm \citep{salvatier_probabilistic_2016}. We follow a prescription similar to \cite{kanodia_toi-5205b_2023}, where we allow eccentricity to float, and include a Gaussian Process (GP) kernel for the \tess~ photometry to detrend out instrument systematics. Similar to previous works, we also include a dilution term to correct the TESS photometry based on the ground-based transits.

We performed joint fitting of the RV timeseries and the photometric curves using \texttt{exoplanet} \citep{foreman-mackey_exoplanet-devexoplanet_2021}, a software package that utilizes \texttt{PyMC3}, a Hamiltonian Monte Carlo (HMC) posterior sampling algorithm \citep{salvatier_probabilistic_2016}.  HMC is a Markov chain Monte Carlo method that uses first-order gradients to avoid random walk behavior, that is further improved by the implementation of the No U Turn Sampler (NUTS) which reduces sensitivity to user specified parameters \citep{hoffman_no-u-turn_2014}.  We use a Keplerian orbit to model the RVs, leaving the eccentricity as a free parameter.  A linear trend and RV offsets for each instrument were fit to the data to account for long-term changes from both astrophysical and instrumental causes.

We modeled the transits using \texttt{starry} \citep{luger_starry_2019, agol_analytic_2020}, which uses the analytic formulae derived in \cite{mandel_analytic_2002} to compute the light curves based on model stellar atmospheres and relies on separate quadratic limb darkening parameters for each filter. During the joint fit with all RV and photometric instruments, we fit each phased transit with separate limb-darkening coefficients.  A jitter term was fit for each data set as a white noise model that was then added in quadrature to the uncertainty of the data sets.  We used \texttt{celerite2} \citep{foreman-mackey_fast_2017, foreman-mackey_scalable_2018} to include a Gaussian Process (GP) kernel in the likelihood function of the TESS data to model the quasi-periodic signal.

A dilution term, $D_{\rm{TESS}}$ is included for each of the two TESS transits as the larger pixel scale can lead to contamination from background stars that alter the transit signal.  We assume the ground based transits do not experience flux contamination due to their higher spatial resolution and use them to correct the TESS photometry.  We fit a separate term for each sector due to variations in the placement of the target and background stars on the detector pixels.  A linear trend was seen in the residuals of the $g^\prime$ band transit from the LCOGT, which was fit out prior to inclusion in the joint fit.

\begin{figure*}[h!]
    \centering
    \includegraphics[scale=0.43]{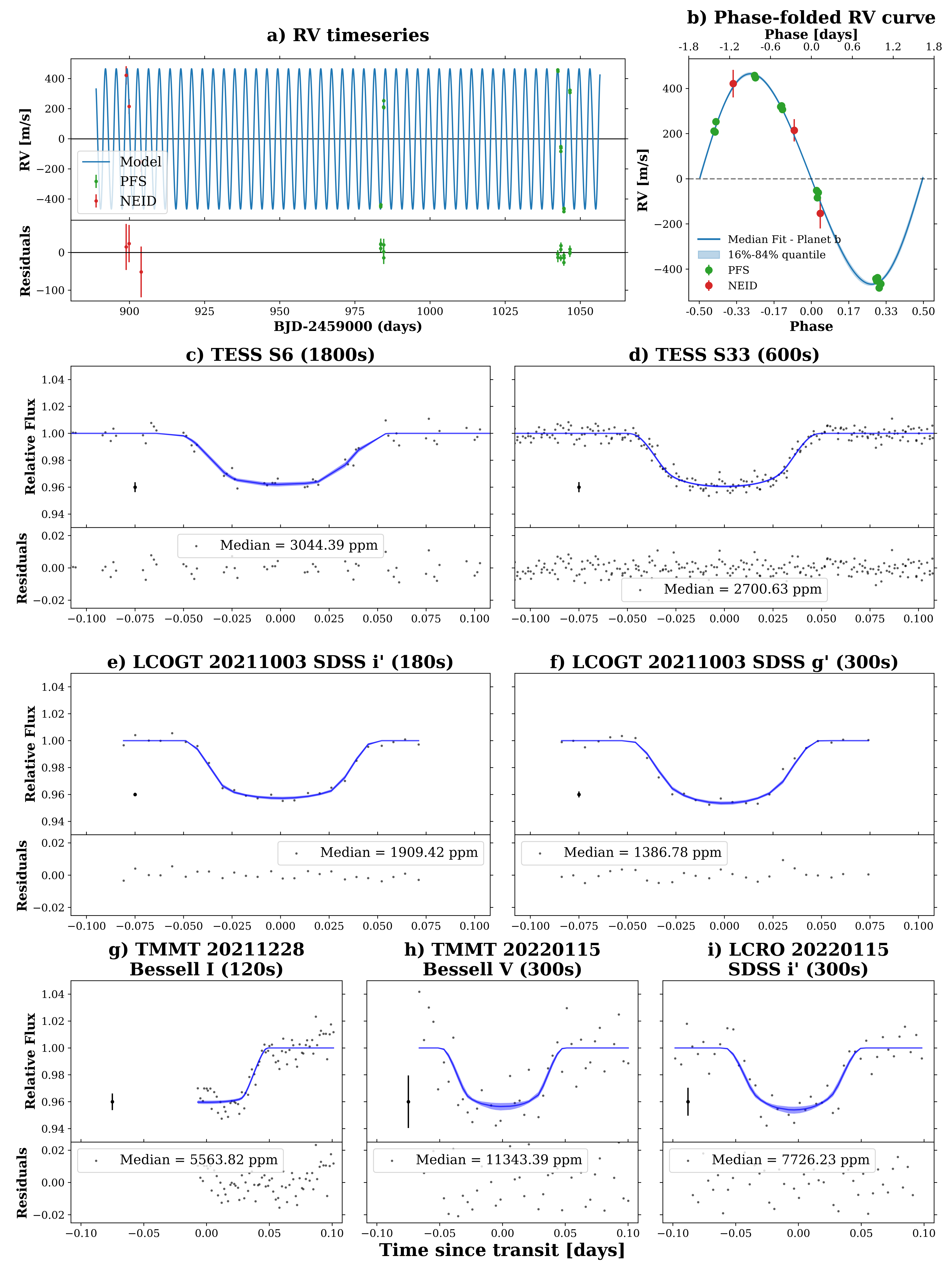}
    \caption{Figure including all photometric and RV observations used in the analysis of TOI-4201~b. (a) Timeseries of RV observations of TOI-4201~b with NEID (red) and PFS (green). The best fitting model from the joint fit is plotted in blue and residuals after subtracting the fit are included below. (b) NEID and PFS observations phase-folded on the best fit orbital period from the joint fit, with the model in blue and 16-84\% confidence levels in light blue. (c)-(i) Photometric observations of TOI-4201~b; in all plots, the detrended data are in grey and the model and 16-18\% confidence levels are in blue. In each figure, the point at x=-0.075 represents the median uncertainty in the photometric data.}
    \label{fig:orbitfig}
\end{figure*}

The final derived planet parameters from the joint fit are included in Table \ref{tab:planetprop} and the phased transits and RVs are shown in Fig \ref{fig:orbitfig}. While we estimate a non-zero eccentricity, we note that this is dominated by the PFS observations as evinced by the low RV jitter compared to NEID. Previous work on HATS-6~b also derived a non-zero eccentricity using PFS which was inconsistent with stellar parameters and it was suggested there may be an underestimation of the formal RV errors when it comes to faint targets \citep{hartman_hats-6b_2015}. While we formally adopt the eccentric fit here in the interest of completeness, we caution against over-interpretation of this tentative eccentricity detection.

\begin{deluxetable*}{llc}
\tablecaption{Planetary parameters for the TOI-4201 System. \label{tab:planetprop}}
\tablehead{\colhead{~~~Parameter} &
\colhead{Units} &
\colhead{Value$^a$} 
}
\startdata
\sidehead{Orbital Parameters:}
~~~Orbital Period\dotfill & $P$ (days) \dotfill & 3.5819210$^{+0.0000038}_{-0.0000039}$\\
~~~Eccentricity\dotfill & $e$ \dotfill & 0.069$^{+0.020}_{-0.021}$\\
~~~Argument of Periastron\dotfill & $\omega$ (radians) \dotfill & -1.61$^{+0.12}_{-0.20}$ \\
~~~Semi-amplitude Velocity\dotfill & $K$ (\ms{})\dotfill &
466.7$^{+5.6}_{-5.7}$\\
~~~Systemic Velocity$^b$\dotfill & $\gamma_{\mathrm{NEID}}$ (\ms{})\dotfill & -297$^{+61}_{-63}$\\
~~~ & $\gamma_{\mathrm{PFS}}$ (\ms{})\dotfill & 51.8$^{+4.7}_{-4.6}$\\
~~~RV trend\dotfill & $dv/dt$ (\ms{} yr$^{-1}$)   & -0.8$\pm5.0$ \\ 
~~~RV jitter\dotfill & $\sigma_{\mathrm{NEID}}$ (\ms{})\dotfill & 69$^{+154}_{-51}$\\
~~~ & $\sigma_{\mathrm{PFS}}$ (\ms{})\dotfill & 12.9$^{+5.4}_{-4.5}$\\
\sidehead{Transit Parameters:}
~~~Transit Midpoint \dotfill & $T_C$ (BJD\textsubscript{TDB})\dotfill & 2459205.255209$^{+0.000293}_{-0.000296}$\\
~~~Scaled Radius\dotfill & $R_{p}/R_{*}$ \dotfill & 
0.1963$^{+0.0030}_{-0.0032}$\\
~~~Scaled Semi-major Axis\dotfill & $a/R_{*}$ \dotfill & 13.83$^{+0.39}_{-0.36}$\\
~~~Orbital Inclination\dotfill & $i$ (degrees)\dotfill & 88.09$^{+0.23}_{-0.20}$\\
~~~Transit Duration\dotfill & $T_{14}$ (days)\dotfill & 0.0900$^{+0.0020}_{-0.0018}$\\
~~~Photometric Jitter$^c$ \dotfill & $\sigma_{\mathrm{TESS~S6}}$ (ppm)\dotfill & 2870$^{+170}_{-180}$\\ 
~~~ & $\sigma_{\mathrm{TESS~S33}}$ (ppm)\dotfill & 1420$^{+130}_{-150}$\\ 
~~~ & $\sigma_{\mathrm{LCO~20211003~i}}$ (ppm)\dotfill & 2130$^{+510}_{-430}$\\ 
~~~ & $\sigma_{\mathrm{LCO~20211003~g}}$ (ppm)\dotfill & 1800$^{+1100}_{-1800}$\\ 
~~~ & $\sigma_{\mathrm{TMMT~20211228}}$ (ppm)\dotfill & 2000$^{+2500}_{-2000}$\\ 
~~~ & $\sigma_{\mathrm{TMMT~20220115}}$ (ppm)\dotfill & 80$^{+1600}_{-70}$\\ 
~~~ & $\sigma_{\mathrm{LCRO~20220115}}$ (ppm)\dotfill & 70$^{+1100}_{-60}$\\ 
~~~Dilution$^{de}$ \dotfill & $D_{\mathrm{TESS~S6}}$ \dotfill & 0.909$^{+0.039}_{-0.038}$\\
~~~ & $D_{\mathrm{TESS~S33}}$ \dotfill & 0.876$^{+0.025}_{-0.024}$\\ 
\sidehead{Planetary Parameters:}
~~~Mass\dotfill & $M_{p}$ (M$_\oplus$)\dotfill &  823$^{+21}_{-20}$\\
~~~ & $M_{p}$ ($M_J$)\dotfill &  2.589$^{+0.066}_{-0.063}$\\
~~~Radius\dotfill & $R_{p}$  (R$_\oplus$) \dotfill& 13.11$^{+0.45}_{-0.46}$\\
~~~ & $R_{p}$  ($R_J$) \dotfill& 1.169$^{+0.040}_{-0.041}$\\
~~~Density\dotfill & $\rho_{p}$ (\gcmcubed{})\dotfill & 2.01$^{+0.23}_{-0.20}$\\
~~~Semi-major Axis\dotfill & $a$ (AU) \dotfill & 0.03939$\pm{0.00041}$\\
~~~Average Incident Flux$^f$\dotfill & $\langle F \rangle$ (\unit{10^5\ W/m^2})\dotfill &  0.697$\pm$0.053\\
~~~Planetary Insolation & $S$ (S$_\oplus$)\dotfill &  51.2$\pm3.9$\\
~~~Equilibrium Temperature$^g$ \dotfill & $T_{\mathrm{eq}}$ (K)\dotfill & 745$\pm14$\\
\enddata
\tablenotetext{a}{The reported values refer to the 16-50-84\% percentile of the posteriors.}
\tablenotetext{b}{In addition to the "Absolute RV" from \autoref{tab:stellarparam}.}
\tablenotetext{c}{Jitter (per observation) added in quadrature to photometric instrument error.}
\tablenotetext{d}{Dilution due to presence of background stars in \tess{} aperture, not accounted for in the \texttt{eleanor} flux.}
\tablenotetext{e}{We use a Solar flux constant = 1360.8 W/m$^2$, to convert insolation to incident flux.}
\tablenotetext{f}{We assume the planet to be a black body with zero albedo and perfect energy redistribution to estimate the equilibrium temperature. }
\normalsize
\end{deluxetable*}

\section{Discussion}\label{sec:discussion}

\subsection{Placing TOI-4201~b in context}
\begin{figure*}
    \centering
    \includegraphics[scale=0.7]{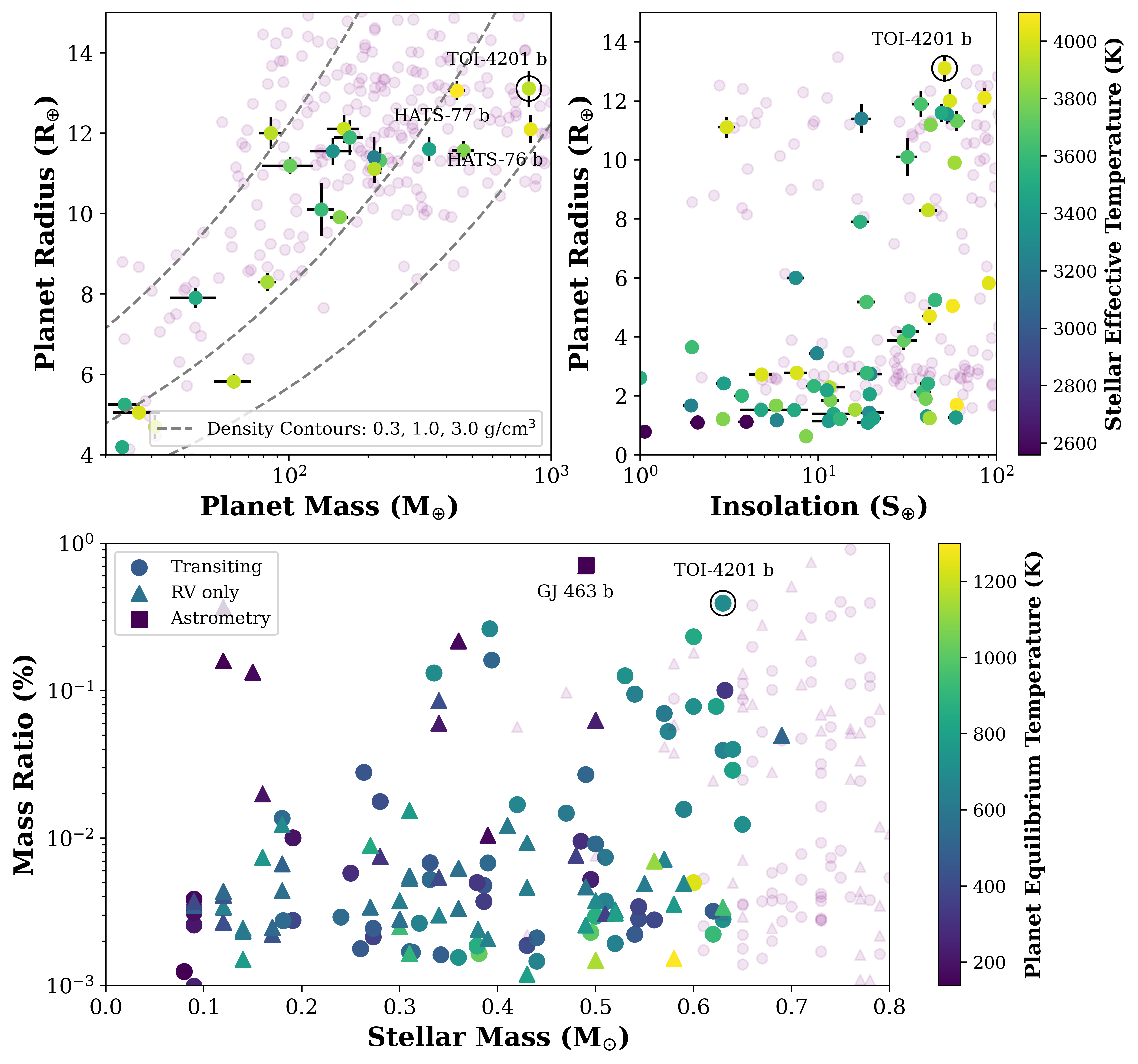}
    \caption{\textbf{Upper left:} We include TOI-4201~b (circled in black) on a mass-radius plane, with planets colored by host star temperature. We include planets around FGK stars in the background and grey contours indicate bulk densities of 0.3, 1.0, and 3.0\,g/cm$^3$ (left to right). \textbf{Upper right:} TOI-4201~b in an insolation-radius plane for the same sample of planets. \textbf{Lower:} Planet to star mass ratio vs. host star mass, colored by equilibrium temperature. Planets with a true mass measurement from transit observations are represented by circular points and planets with a true mass measurement from astrometry are squares, while triangles are minimum masses (RV only). Around M-dwarfs, TOI-4201~b has the highest mass ratio for transiting planets and the planet with the highest mass ratio overall is GJ 463~b \citep{endl_jupiter_2022,sozzetti_dynamical_2023}.}
    \label{fig:parameterspace}
\end{figure*}

TOI-4201~b is a massive Jovian planet with a radius of 1.17$\pm0.04$ $R_J$, mass 2.59$^{+0.07}_{-0.06}$ $M_J$, and density 2.0$\pm0.2$\,g\,cm$^{-3}$.  Orbiting a metal-rich M-dwarf host star (M$_\star$ = 0.63$\pm0.02$\,M$_\odot$), it joins a small but growing number of GEMS with precise masses and radii. Additionally, at a mass ratio of $\sim0.39$\%, it is just below the new limit of 0.4\% for exoplanets that has been proposed by the IAU \citep{Lecavelier_IAUdef_2022}. In \autoref{fig:parameterspace}, we plot TOI-4201~b together with all known transiting giant planets (R $>$ 4\,R$_\oplus$) around M-dwarfs, up to T$_{\rm{eff}}< $4100\,K to account for host stars on the late K/early M border in the first panel. Subsequent panels use a cutoff of T$_{\rm{eff}}< $4000\,K.

\begin{figure}[h!]
    \centering
    \includegraphics[width=0.48\textwidth]{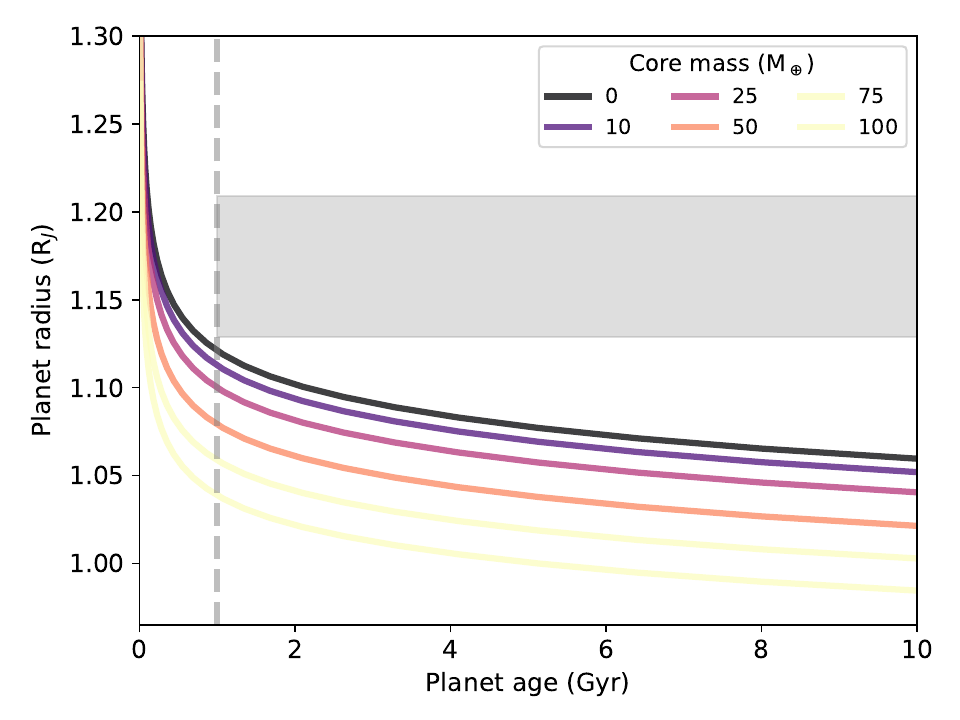}
    \caption{The radius evolution of a planet with the same mass and equilibrium temperature as TOI-4201~b assuming different core (heavy-element) masses, following the model described in \cite{muller_synthetic_2021}. The grey shaded region represents the area of this parameter space in which our analysis places TOI-4201~b, a region substantially above the radius expected for a planet composed purely of H/He. As there are no indications this is a particularly young system, we include a dashed line at 1\,Gyr to suggest a lower limit on the age of the planet.}
    \label{fig:inflateradius}
\end{figure}

As shown in \autoref{fig:parameterspace}, the closest grouping in mass-radius space to TOI-4201~b consists of HATS-76~b and HATS-77~b \citep{jordan_hats-74ab_2022}, both of which orbit stars that are on the edge between late K and early M-dwarfs. Of particular note is the similarity in radius between TOI-4201~b and HATS-77~b; despite both planets orbiting old inactive stars\footnote{While we do not have a precise age constraint on TOI-4201, the photometry and spectroscopic observations do not contain activity signatures common to young stars, suggesting an old system.}, and neither planet experiencing sufficient irradiation to inflate the radius \citep{demory_inflation_2011}, both have somewhat larger radii than models would suggest \citep{mordasini_extrasolar_2012}. In order to characterize the interior of TOI-4201~b and quantify the degree of inflation, we use the giant planet evolution models from \cite{muller_synthetic_2021} and calculate the cooling for different possible heavy-element masses. Using the derived planetary parameters, these models suggest that the planet is inflated by $\sim$ 5--10\% beyond what would be expected for a planet with a pure H/He composition as shown in \autoref{fig:inflateradius}.

Previous work has found that the assumptions underlying interior models of giant planets can cause variations in the derived radius on the order of a few percent \citep[for a review, see][]{Muller_giantplanet)_review_2023}. The main culprits appear to be uncertainties in the H-He equations of state \citep{muller_theoretical_2020, Howard_nonidealHHe_2023} and the opacity \citep{muller_theoretical_2020}. Additionally, current evolution models of giant exoplanets assume that their interiors are fully convective. However, there is evidence that Jupiter and Saturn have regions that are not fully convective today due to composition gradients \citep{debras_jupiterconvection_2021, Mankovich_saturn_convection_2021}. This could suggest that giant planets do not cool predominantly by large-scale convection, and therefore their interiors may be hotter, leading to inflated radii of up to about 10\% past 1 Gyr \citep{kurokawa_radiusinflation_2015}. However, it is currently unclear whether these composition gradients could be sustained over a few Gyr \citep{muller_challenge_2020}. To explain the apparent inflation of TOI-4201 b, either its cooling must be slowed by the aforementioned mechanisms, or there must be another process heating the interior. However, a detailed investigation of this would require next-generation evolution models, and is beyond the scope of this work.

\subsection{Planet formation and migration}

Core accretion is a model of planet formation by which small solid particles coagulate and gradually grow to planetary embryos either through pebble or planetesimal formation. These embryos can be massive enough to trigger runaway gas accretion and allow for planets to retain large H/He dominated atmospheres \citep{pollack_formation_1996}. While this is the favored model for close in planets, the decreased solid mass available \citep{andrews_mass_2013, pascucci_steeper_2016} and the increased Keplerian orbital timescales \citep{laughlin_core_2004} around M-dwarfs make reaching runaway accretion and forming giant planets a challenge.

We use a basic mass budget framework to determine the possibility of TOI-4201~b forming through core-accretion. TOI-4201~b seems to be inflated beyond what current theoretical models for giant planets support, so we are unable to use available planetary interior and evolution models such as \texttt{planetsynth} meaningfully to find a heavy-element content or bulk metallicity \citep{muller_synthetic_2021}. Instead, we make a lower-bound estimate of the heavy-element content by assuming the planet will have approximately the same metallicity as its host star. The composition of the Sun is approximately 1.39$\pm0.06$\% metals by mass \citep{Asplund_solar_metal_content_2021}. Combining this with a metallicity of 0.30$\pm0.15$\,dex for TOI-4201 yields an estimated heavy-element mass of 23$\pm8$\,M$_{\oplus}$ for TOI-4201~b. Conservatively, if we assume 10\% formation efficiency \citep{liu_growth_2019} this would require a minimum dust mass of $\sim 230\pm80$\,M$_\oplus$ to have been present for planet formation in the Class II disk. In the Lupus association, the dust available in Class II disks around M dwarfs ranges from 1--50\,M$_{\oplus}$ \citep{manara_demographics_2022}, suggesting it is unlikely there would be sufficient material for TOI-4201~b to form through core accretion in these disks.


There are caveats to this simple argument; the efficiency of the core accretion mechanism is poorly constrained and our understanding of the dust masses contained in disks when planet formation begins is incomplete. In our argument we have assumed a 10\% efficiency, but simulations have shown this value is strongly dependent on both the turbulence of the disk and the fragmentation velocity of individual particles. Under reasonable disk conditions this efficiency may range from $\sim$1\% up to 40\% \citep{Guillot_filtering_and_processing_2014,Chachan_small_planets_2023}. Ring structures in the protoplanetary disk may also increase formation efficiency and leave behind wide planetesimal belts with distinct profiles exterior to the planets in a system \citep{Jiang_planets_in_rings_2023}. While a debris belt is unlikely to be detected around an older star like TOI-4201, the detection of younger GEMS may present an opportunity to look for this profile as a clue towards the formation environment of similar objects.

The dust mass of a disk is typically derived by assuming blackbody emission and using single wavelength continuum flux measurements, often in the millimeter/sub-millimeter regime, combined with an assumption that the emission is optically thin \citep{hildebrand_determination_1983}. Recent results have shown this is likely to underestimate the solid mass in the disk; full radiative transfer modeling across multiple wavelengths has shown continuum emission is likely optically thick \citep{arnaud_VLA1623_mwvl_2022, xin_measuring_2023} and dust masses derived by SED fitting are greater than the analytical estimate by a factor of 1.5 -- 5 \citep{rilinger_determining_2023}. One possibility is that a significant amount of the dust mass is contained in larger bodies to which millimeter observations are not sensitive \citep{najita_mass_2014}. The exact amount contained in such bodies is unknown, but the upcoming Next Generation Very Large Array (ngVLA) will be sensitive to cm-sized grains and its higher resolution will allow us to probe disk substructure caused by low mass planets \citep{selina_ngVLA_2018, Selina_ngVLA_2022}.

Due to the gaps in our understanding, we cannot completely rule out core accretion as a formation mechanism for TOI-4201~b. However, if we assume a substantial amount of the dust in a Class II disk is contained in larger bodies, this could be indicative of planet formation beginning earlier in the lifetime of the disk \citep{Tsukamoto_apparent_diskmass_2017}. Formation beginning in the Class I/0 phase can also decrease the accretion timescale, forming a core large enough to trigger runaway gas accretion within 0.5\,Myr \citep{Tanaka_pebble_accretion_classI/0_2019}. Ongoing work with ALMA is looking for substructures in these early disks to better constrain when planet formation begins \citep{Ohashi_eDisk_2023}.

If planet formation efficiency is $<$10\% then gravitational instability, whereby instabilities in massive circumstellar disks allow for the material to directly collapse into giant planet cores on short timescales \citep[$\sim$10$^3$ -- 10$^4$ years after the disk becomes unstable][]{boss_gravitationalinstability_1997}, could be a viable alternative formation mechanism for TOI-4201~b. While the hope of differentiating between the two mechanisms and therefore gaining insight into formation efficiency through atmospheric spectroscopy is desirable \citep{hobbs_molecular_2022},  the complex and uncertain nature of the models makes any individual results more qualitative than quantitative \citep{molliere_interpreting_2022}. Atmospheric spectroscopy of GEMS is now beginning with JWST (and ARIEL in the future), and may help shed light on lines of attack, or lack thereof, to this problem.

\section{Summary}\label{sec:conclusion}
We present the discovery of TOI-4201~b, a Jovian exoplanet with an inflated radius orbiting an early M dwarf.  The planet was first identified from TESS photometry and follow up observations consisting of ground based photometry, RVs, and speckle imaging constrained the orbital parameters and allowed for characterization of the planet.

The TOI-4201~b mass ratio of $\sim$0.39\% is one of the highest known for transiting planets around M dwarfs.  Assuming a 10\% formation efficiency \citep{liu_growth_2019} and a stellar/sub-stellar atmospheric metallicity and the corresponding planetary heavy-element content of $\sim$20\,M$_\oplus$ would require a disk with a dust mass of $\sim$200\,M$_\oplus$. Better estimates of disk dust mass demonstrate the most massive disks may reach this threshold, but most remain below the threshold. The existence of TOI-4201~b is suggestive of planet formation beginning before the Class II disk phase.

\newpage
\section{Acknowledgements}

Data presented herein were obtained at the WIYN Observatory from telescope time allocated to NN-EXPLORE through the scientific partnership of the National Aeronautics and Space Administration, the National Science Foundation, and NOIRLab. This work was supported by a NASA WIYN PI Data Award, administered by the NASA Exoplanet Science Institute. These results are based on observations obtained with NEID on the WIYN 3.5 m telescope at KPNO, NSF's NOIRLab under proposal 2022B-785506 (PI: S. Kanodia), managed by the Association of Universities for Research in Astronomy (AURA) under a cooperative agreement with the NSF. This work was performed for the Jet Propulsion Laboratory, California Institute of Technology, sponsored by the United States Government under the Prime Contract 80NM0018D0004 between Caltech and NASA.

WIYN is a joint facility of the University of Wisconsin-Madison, Indiana University, NSF's NOIRLab, the Pennsylvania State University, Purdue University, University of California-Irvine, and the University of Missouri. 

The authors are honored to be permitted to conduct astronomical research on Iolkam Du'ag (Kitt Peak), a mountain with particular significance to the Tohono O'odham. Data presented herein were obtained at the WIYN Observatory from telescope time allocated to NN-EXPLORE through the scientific partnership of NASA, the NSF, and NOIRLab.

Some of the observations in this paper made use of the NN-EXPLORE Exoplanet and Stellar Speckle Imager (NESSI). NESSI was funded by the NASA Exoplanet Exploration Program and the NASA Ames Research Center. NESSI was built at the Ames Research Center by Steve B. Howell, Nic Scott, Elliott P. Horch, and Emmett Quigley.

This work has made use of data from the European Space Agency (ESA) mission Gaia (\url{https://www.cosmos.esa.int/gaia}), processed by the Gaia Data Processing and Analysis Consortium (DPAC, \url{https://www.cosmos.esa.int/web/gaia/dpac/consortium}). Funding for the DPAC has been provided by national institutions, in particular the institutions participating in the Gaia Multilateral Agreement.

Some of the observations in this paper were obtained with the Samuel Oschin Telescope 48-inch and the 60-inch Telescope at the Palomar Observatory as part of the ZTF project. ZTF is supported by the NSF under Grant No. AST-2034437 and a collaboration including Caltech, IPAC, the Weizmann Institute for Science, the Oskar Klein Center at Stockholm University, the University of Maryland, Deutsches Elektronen-Synchrotron and Humboldt University, the TANGO Consortium of Taiwan, the University of Wisconsin at Milwaukee, Trinity College Dublin, Lawrence Livermore National Laboratories, and IN2P3, France. Operations are conducted by COO, IPAC, and UW.

This work makes use of observations (Proposal ID: KEY2020B-005) from the Sinistro imaging camera on the 1m Dome B telescope at Cerro Tololo Inter-American Observatory, operated by the Las Cumbres Observatory global telescope network (LCOGT).

Computations for this research were performed on the Pennsylvania State University’s Institute for Computational and Data Sciences Advanced CyberInfrastructure (ICDS-ACI).  This content is solely the responsibility of the authors and does not necessarily represent the views of the Institute for Computational and Data Sciences.

The Center for Exoplanets and Habitable Worlds is supported by the Pennsylvania State University, the Eberly College of Science, and the Pennsylvania Space Grant Consortium. 

Some of the data presented in this paper were obtained from MAST at STScI. Support for MAST for non-HST data is provided by the NASA Office of Space Science via grant NNX09AF08G and by other grants and contracts.

This work includes data collected by the TESS mission, which are publicly available from MAST. Funding for the TESS mission is provided by the NASA Science Mission directorate. 

This research made use of the (i) NASA Exoplanet Archive, which is operated by Caltech, under contract with NASA under the Exoplanet Exploration Program, (ii) SIMBAD database, operated at CDS, Strasbourg, France, (iii) NASA's Astrophysics Data System Bibliographic Services, and (iv) data from 2MASS, a joint project of the University of Massachusetts and IPAC at Caltech, funded by NASA and the NSF.

This research has made use of the SIMBAD database, operated at CDS, Strasbourg, France, 
and NASA's Astrophysics Data System Bibliographic Services.

This research has made use of the Exoplanet Follow-up Observation Program website, which is operated by the California Institute of Technology, under contract with the National Aeronautics and Space Administration under the Exoplanet Exploration Program

CIC acknowledges support by NASA Headquarters through an appointment to the NASA Postdoctoral Program at the Goddard Space Flight Center, administered by USRA through a contract with NASA.

GS acknowledges support provided by NASA through the NASA Hubble Fellowship grant HST-HF2-51519.001-A awarded by the Space Telescope Science Institute, which is operated by the Association of Universities for Research in Astronomy, Inc., for NASA, under contract NAS5-26555.

\facilities{\gaia{}, WIYN (NEID), WIYN (NESSI), Magellan (PFS), \tess{}, Exoplanet Archive}
\software{
\texttt{ArviZ} \citep{kumar_arviz_2019}, 
AstroImageJ \citep{collins_astroimagej_2017}, 
\texttt{astroquery} \citep{ginsburg_astroquery_2019}, 
\texttt{astropy} \citep{robitaille_astropy_2013, astropy_collaboration_astropy_2018},
\texttt{barycorrpy} \citep{kanodia_python_2018}, 
\texttt{celerite2} \citep{foreman-mackey_fast_2017, foreman-mackey_scalable_2018}
\texttt{exoplanet} \citep{foreman-mackey_exoplanet-devexoplanet_2021, foreman-mackey_exoplanet_2021},
\texttt{ipython} \citep{perez_ipython_2007},
\texttt{lightkurve} \citep{lightkurve_collaboration_lightkurve_2018},
\texttt{matplotlib} \citep{hunter_matplotlib_2007},
\texttt{numpy} \citep{oliphant_numpy_2006},
\texttt{pandas} \citep{mckinney_data_2010},
\texttt{pyastrotools} \citep{kanodia_shubham_2023_7685628}
\texttt{PyMC3} \citep{salvatier_probabilistic_2016},
\texttt{scipy} \citep{oliphant_python_2007, virtanen_scipy_2020},
\texttt{SERVAL} \citep{zechmeister_spectrum_2018},
\texttt{starry} \citep{luger_starry_2019, agol_analytic_2020},
\texttt{Theano} \citep{the_theano_development_team_theano_2016},
\texttt{planetsynth} \citep{muller_synthetic_2021}.}

\bibliography{references,manualrefs}

\begin{thebibliography}{}
\expandafter\ifx\csname natexlab\endcsname\relax\def\natexlab#1{#1}\fi
\providecommand{\url}[1]{\href{#1}{#1}}

\bibitem[{Agol {et~al.}(2020)Agol, Luger, \&
  Foreman-Mackey}]{agol_analytic_2020}
Agol, E., Luger, R., \& Foreman-Mackey, D. 2020, The Astronomical Journal, 159,
  123.
\newblock \url{https://ui.adsabs.harvard.edu/abs/2020AJ....159..123A}

\bibitem[{Anders {et~al.}(2022)Anders, Khalatyan, Queiroz, Chiappini, Ardèvol,
  Casamiquela, Figueras, Jiménez-Arranz, Jordi, Monguió, Romero-Gómez,
  Altamirano, Antoja, Assaad, Cantat-Gaudin, Castro-Ginard, Enke, Girardi,
  Guiglion, Khan, Luri, Miglio, Minchev, Ramos, Santiago, \&
  Steinmetz}]{anders_photo-astrometric_2022}
Anders, F., Khalatyan, A., Queiroz, A. B.~A., {et~al.} 2022, Astronomy \&amp;
  Astrophysics, Volume 658, id.A91,
  {\textless}NUMPAGES{\textgreater}27{\textless}/NUMPAGES{\textgreater} pp.,
  658, A91.
\newblock
  \url{https://ui.adsabs.harvard.edu/abs/2022A%26A...658A..91A/abstract}

\bibitem[{Andrews {et~al.}(2013)Andrews, Rosenfeld, Kraus, \&
  Wilner}]{andrews_mass_2013}
Andrews, S.~M., Rosenfeld, K.~A., Kraus, A.~L., \& Wilner, D.~J. 2013, The
  Astrophysical Journal, 771, 129, aDS Bibcode: 2013ApJ...771..129A.
\newblock \url{https://ui.adsabs.harvard.edu/abs/2013ApJ...771..129A}

\bibitem[{Ansdell {et~al.}(2016)Ansdell, Williams, van~der Marel, Carpenter,
  Guidi, Hogerheijde, Mathews, Manara, Miotello, Natta, Oliveira, Tazzari,
  Testi, van Dishoeck, \& van Terwisga}]{ansdell_alma_2016}
Ansdell, M., Williams, J.~P., van~der Marel, N., {et~al.} 2016, The
  Astrophysical Journal, 828, 46, aDS Bibcode: 2016ApJ...828...46A.
\newblock \url{https://ui.adsabs.harvard.edu/abs/2016ApJ...828...46A}

\bibitem[{{Asplund} {et~al.}(2021){Asplund}, {Amarsi}, \&
  {Grevesse}}]{Asplund_solar_metal_content_2021}
{Asplund}, M., {Amarsi}, A.~M., \& {Grevesse}, N. 2021, \aap, 653, A141

\bibitem[{{Astropy Collaboration} {et~al.}(2018){Astropy Collaboration},
  Price-Whelan, Sipőcz, Günther, Lim, Crawford, Conseil, Shupe, Craig,
  Dencheva, Ginsburg, VanderPlas, Bradley, Pérez-Suárez, de~Val-Borro,
  Aldcroft, Cruz, Robitaille, Tollerud, Ardelean, Babej, Bach, Bachetti,
  Bakanov, Bamford, Barentsen, Barmby, Baumbach, Berry, Biscani, Boquien,
  Bostroem, Bouma, Brammer, Bray, Breytenbach, Buddelmeijer, Burke, Calderone,
  Cano~Rodríguez, Cara, Cardoso, Cheedella, Copin, Corrales, Crichton,
  D'Avella, Deil, Depagne, Dietrich, Donath, Droettboom, Earl, Erben, Fabbro,
  Ferreira, Finethy, Fox, Garrison, Gibbons, Goldstein, Gommers, Greco,
  Greenfield, Groener, Grollier, Hagen, Hirst, Homeier, Horton, Hosseinzadeh,
  Hu, Hunkeler, Ivezić, Jain, Jenness, Kanarek, Kendrew, Kern, Kerzendorf,
  Khvalko, King, Kirkby, Kulkarni, Kumar, Lee, Lenz, Littlefair, Ma, Macleod,
  Mastropietro, McCully, Montagnac, Morris, Mueller, Mumford, Muna, Murphy,
  Nelson, Nguyen, Ninan, Nöthe, Ogaz, Oh, Parejko, Parley, Pascual, Patil,
  Patil, Plunkett, Prochaska, Rastogi, Reddy~Janga, Sabater, Sakurikar,
  Seifert, Sherbert, Sherwood-Taylor, Shih, Sick, Silbiger, Singanamalla,
  Singer, Sladen, Sooley, Sornarajah, Streicher, Teuben, Thomas, Tremblay,
  Turner, Terrón, van Kerkwijk, de~la Vega, Watkins, Weaver, Whitmore,
  Woillez, Zabalza, \& {Astropy
  Contributors}}]{astropy_collaboration_astropy_2018}
{Astropy Collaboration}, Price-Whelan, A.~M., Sipőcz, B.~M., {et~al.} 2018,
  The Astronomical Journal, 156, 123.
\newblock \url{https://ui.adsabs.harvard.edu/abs/2018AJ....156..123A}

\bibitem[{Bailer-Jones {et~al.}(2021)Bailer-Jones, Rybizki, Fouesneau,
  Demleitner, \& Andrae}]{bailer-jones_estimating_2021}
Bailer-Jones, C. A.~L., Rybizki, J., Fouesneau, M., Demleitner, M., \& Andrae,
  R. 2021, The Astronomical Journal, 161, 147.
\newblock \url{http://adsabs.harvard.edu/abs/2021AJ....161..147B}

\bibitem[{Bakos {et~al.}(2020)Bakos, Bayliss, Bento, Bhatti, Brahm, Csubry,
  Espinoza, Hartman, Henning, Jordán, Mancini, Penev, Rabus, Sarkis, Suc,
  de~Val-Borro, Zhou, Butler, Crane, Durkan, Shectman, Kim, Lázár, Papp,
  Sári, Ricker, Vanderspek, Latham, Seager, Winn, Jenkins, Chacon, Fűrész,
  Goeke, Li, Quinn, Quintana, Tenenbaum, Teske, Vezie, Yu, Stockdale, Evans, \&
  Relles}]{bakos_hats-71b_2020}
Bakos, G.~A., Bayliss, D., Bento, J., {et~al.} 2020, The Astronomical Journal,
  159, 267, aDS Bibcode: 2020AJ....159..267B.
\newblock \url{https://ui.adsabs.harvard.edu/abs/2020AJ....159..267B}

\bibitem[{Bensby {et~al.}(2014)Bensby, Feltzing, \&
  Oey}]{bensby_exploring_2014}
Bensby, T., Feltzing, S., \& Oey, M.~S. 2014, Astronomy and Astrophysics, 562,
  A71.
\newblock \url{http://adsabs.harvard.edu/abs/2014A%26A...562A..71B}

\bibitem[{Birky {et~al.}(2020)Birky, Hogg, Mann, \&
  Burgasser}]{birky_temperatures_2020}
Birky, J., Hogg, D.~W., Mann, A.~W., \& Burgasser, A. 2020, The Astrophysical
  Journal, 892, 31, aDS Bibcode: 2020ApJ...892...31B.
\newblock \url{https://ui.adsabs.harvard.edu/abs/2020ApJ...892...31B}

\bibitem[{{Boss}(1997)}]{boss_gravitationalinstability_1997}
{Boss}, A.~P. 1997, Science, 276, 1836

\bibitem[{Bovy(2015)}]{bovy_galpy_2015}
Bovy, J. 2015, The Astrophysical Journal Supplement Series, 216, 29.
\newblock \url{http://adsabs.harvard.edu/abs/2015ApJS..216...29B}

\bibitem[{{Brown} {et~al.}(2013){Brown}, {Baliber}, {Bianco}, {Bowman},
  {Burleson}, {Conway}, {Crellin}, {Depagne}, {De Vera}, {Dilday}, {Dragomir},
  {Dubberley}, {Eastman}, {Elphick}, {Falarski}, {Foale}, {Ford}, {Fulton},
  {Garza}, {Gomez}, {Graham}, {Greene}, {Haldeman}, {Hawkins}, {Haworth},
  {Haynes}, {Hidas}, {Hjelstrom}, {Howell}, {Hygelund}, {Lister}, {Lobdill},
  {Martinez}, {Mullins}, {Norbury}, {Parrent}, {Paulson}, {Petry}, {Pickles},
  {Posner}, {Rosing}, {Ross}, {Sand}, {Saunders}, {Shobbrook}, {Shporer},
  {Street}, {Thomas}, {Tsapras}, {Tufts}, {Valenti}, {Vander Horst}, {Walker},
  {White}, \& {Willis}}]{brown_2013_LCOGT}
{Brown}, T.~M., {Baliber}, N., {Bianco}, F.~B., {et~al.} 2013, \pasp, 125, 1031

\bibitem[{Butler {et~al.}(1996)Butler, Marcy, Williams, McCarthy, Dosanjh, \&
  Vogt}]{butler_attaining_1996}
Butler, R.~P., Marcy, G.~W., Williams, E., {et~al.} 1996, Publications of the
  Astronomical Society of the Pacific, 108, 500.
\newblock \url{http://adsabs.harvard.edu/abs/1996PASP..108..500B}

\bibitem[{{Chachan} \& {Lee}(2023)}]{Chachan_small_planets_2023}
{Chachan}, Y., \& {Lee}, E.~J. 2023, arXiv e-prints, arXiv:2305.00803

\bibitem[{Collins {et~al.}(2017)Collins, Kielkopf, Stassun, \&
  Hessman}]{collins_astroimagej_2017}
Collins, K.~A., Kielkopf, J.~F., Stassun, K.~G., \& Hessman, F.~V. 2017, The
  Astronomical Journal, 153, 77.
\newblock \url{http://adsabs.harvard.edu/abs/2017AJ....153...77C}

\bibitem[{Crane {et~al.}(2006)Crane, Shectman, \& Butler}]{crane_carnegie_2006}
Crane, J.~D., Shectman, S.~A., \& Butler, R.~P. 2006, 6269, 626931, conference
  Name: Society of Photo-Optical Instrumentation Engineers (SPIE) Conference
  Series ADS Bibcode: 2006SPIE.6269E..31C.
\newblock \url{https://ui.adsabs.harvard.edu/abs/2006SPIE.6269E..31C}

\bibitem[{Crane {et~al.}(2010)Crane, Shectman, Butler, Thompson, Birk, Jones,
  \& Burley}]{crane_carnegie_2010}
Crane, J.~D., Shectman, S.~A., Butler, R.~P., {et~al.} 2010, 7735, 773553,
  conference Name: Ground-based and Airborne Instrumentation for Astronomy III
  ADS Bibcode: 2010SPIE.7735E..53C.
\newblock \url{https://ui.adsabs.harvard.edu/abs/2010SPIE.7735E..53C}

\bibitem[{Crane {et~al.}(2008)Crane, Shectman, Butler, Thompson, \&
  Burley}]{crane_carnegie_2008}
Crane, J.~D., Shectman, S.~A., Butler, R.~P., Thompson, I.~B., \& Burley, G.~S.
  2008, 7014, 701479, conference Name: Ground-based and Airborne
  Instrumentation for Astronomy II ADS Bibcode: 2008SPIE.7014E..79C.
\newblock \url{https://ui.adsabs.harvard.edu/abs/2008SPIE.7014E..79C}

\bibitem[{Cutri {et~al.}(2003)Cutri, Skrutskie, van Dyk, Beichman, Carpenter,
  Chester, Cambresy, Evans, Fowler, Gizis, Howard, Huchra, Jarrett, Kopan,
  Kirkpatrick, Light, Marsh, McCallon, Schneider, Stiening, Sykes, Weinberg,
  Wheaton, Wheelock, \& Zacarias}]{cutri_2mass_2003}
Cutri, R.~M., Skrutskie, M.~F., van Dyk, S., {et~al.} 2003, "The IRSA 2MASS
  All-Sky Point Source Catalog, NASA/IPAC Infrared Science Archive.
  http://irsa.ipac.caltech.edu/applications/Gator/".
\newblock \url{http://adsabs.harvard.edu/abs/2003tmc..book.....C}

\bibitem[{{Debras} {et~al.}(2021){Debras}, {Chabrier}, \&
  {Stevenson}}]{debras_jupiterconvection_2021}
{Debras}, F., {Chabrier}, G., \& {Stevenson}, D.~J. 2021, \apjl, 913, L21

\bibitem[{{Demory} \& {Seager}(2011)}]{demory_inflation_2011}
{Demory}, B.-O., \& {Seager}, S. 2011, \apjs, 197, 12

\bibitem[{Duque-Arribas {et~al.}(2023)Duque-Arribas, Montes, Tabernero,
  Caballero, Gorgas, \& Marfil}]{duque-arribas_photometric_2023}
Duque-Arribas, C., Montes, D., Tabernero, H.~M., {et~al.} 2023, The
  Astrophysical Journal, 944, 106, aDS Bibcode: 2023ApJ...944..106D.
\newblock \url{https://ui.adsabs.harvard.edu/abs/2023ApJ...944..106D}

\bibitem[{Eastman {et~al.}(2019)Eastman, Rodriguez, Agol, Stassun, Beatty,
  Vanderburg, Gaudi, Collins, \& Luger}]{eastman_exofastv2_2019}
Eastman, J.~D., Rodriguez, J.~E., Agol, E., {et~al.} 2019, {EXOFASTv2}: {A}
  public, generalized, publication-quality exoplanet modeling code, Tech. rep.,
  publication Title: arXiv e-prints ADS Bibcode: 2019arXiv190709480E Type:
  article.
\newblock \url{https://ui.adsabs.harvard.edu/abs/2019arXiv190709480E}

\bibitem[{Endl {et~al.}(2022)Endl, Robertson, Cochran, MacQueen, Bowler,
  Franson, Holcomb, Beard, Isaacson, Howard, \& Lubin}]{endl_jupiter_2022}
Endl, M., Robertson, P., Cochran, W.~D., {et~al.} 2022, The Astronomical
  Journal, 164, 238, aDS Bibcode: 2022AJ....164..238E.
\newblock \url{https://ui.adsabs.harvard.edu/abs/2022AJ....164..238E}

\bibitem[{Feinstein {et~al.}(2019)Feinstein, Montet, Foreman-Mackey, Bedell,
  Saunders, Bean, Christiansen, Hedges, Luger, Scolnic, \&
  Cardoso}]{feinstein_eleanor_2019}
Feinstein, A.~D., Montet, B.~T., Foreman-Mackey, D., {et~al.} 2019,
  Publications of the Astronomical Society of the Pacific, 131, 094502.
\newblock \url{https://ui.adsabs.harvard.edu/abs/2019PASP..131i4502F}

\bibitem[{Foreman-Mackey(2018)}]{foreman-mackey_scalable_2018}
Foreman-Mackey, D. 2018, Research Notes of the American Astronomical Society,
  2, 31

\bibitem[{Foreman-Mackey {et~al.}(2017)Foreman-Mackey, Agol, Ambikasaran, \&
  Angus}]{foreman-mackey_fast_2017}
Foreman-Mackey, D., Agol, E., Ambikasaran, S., \& Angus, R. 2017, The
  Astronomical Journal, 154, 220, arXiv: 1703.09710.
\newblock \url{http://arxiv.org/abs/1703.09710}

\bibitem[{Foreman-Mackey {et~al.}(2021{\natexlab{a}})Foreman-Mackey, Savel,
  Luger, Czekala, Agol, Price-Whelan, Gilbert, Brandt, Barclay, \&
  Bouma}]{foreman-mackey_exoplanet-devexoplanet_2021}
Foreman-Mackey, D., Savel, A., Luger, R., {et~al.} 2021{\natexlab{a}},
  exoplanet-dev/exoplanet v0.4.4, , , doi:10.5281/zenodo.1998447.
\newblock \url{https://doi.org/10.5281/zenodo.1998447}

\bibitem[{Foreman-Mackey {et~al.}(2021{\natexlab{b}})Foreman-Mackey, Luger,
  Agol, Barclay, Bouma, Brandt, Czekala, David, Dong, Gilbert, Gordon, Hedges,
  Hey, Morris, Price-Whelan, \& Savel}]{foreman-mackey_exoplanet_2021}
Foreman-Mackey, D., Luger, R., Agol, E., {et~al.} 2021{\natexlab{b}}, The
  Journal of Open Source Software, 6, 3285.
\newblock \url{https://ui.adsabs.harvard.edu/abs/2021JOSS....6.3285F}

\bibitem[{Gagné {et~al.}(2018)Gagné, Mamajek, Malo, Riedel, Rodriguez,
  Lafrenière, Faherty, Roy-Loubier, Pueyo, Robin, \&
  Doyon}]{gagne_banyan_2018}
Gagné, J., Mamajek, E.~E., Malo, L., {et~al.} 2018, The Astrophysical Journal,
  856, 23.
\newblock \url{http://adsabs.harvard.edu/abs/2018ApJ...856...23G}

\bibitem[{{Gaia Collaboration} {et~al.}(2022){Gaia Collaboration}, Arenou,
  Babusiaux, Barstow, Faigler, Jorissen, Kervella, Mazeh, Mowlavi, Panuzzo,
  Sahlmann, Shahaf, Sozzetti, Bauchet, Damerdji, Gavras, Giacobbe, Gosset,
  Halbwachs, Holl, Lattanzi, Leclerc, Morel, Pourbaix, Re~Fiorentin, Sadowski,
  Ségransan, Siopis, Teyssier, Zwitter, Planquart, Brown, Vallenari, Prusti,
  de~Bruijne, Biermann, Creevey, Ducourant, Evans, Eyer, Guerra, Hutton, Jordi,
  Klioner, Lammers, Lindegren, Luri, Mignard, Panem, Randich, Sartoretti,
  Soubiran, Tanga, Walton, Bailer-Jones, Bastian, Drimmel, Jansen, Katz, van
  Leeuwen, Bakker, Cacciari, Castañeda, De~Angeli, Fabricius, Fouesneau,
  Frémat, Galluccio, Guerrier, Heiter, Masana, Messineo, Nicolas,
  Nienartowicz, Pailler, Riclet, Roux, Seabroke, Sordo, Thévenin,
  Gracia-Abril, Portell, Altmann, Andrae, Audard, Bellas-Velidis, Benson,
  Berthier, Blomme, Burgess, Busonero, Busso, Cánovas, Carry, Cellino, Cheek,
  Clementini, Davidson, de~Teodoro, Nuñez~Campos, Delchambre, Dell'Oro,
  Esquej, Fernández-Hernández, Fraile, Garabato, García-Lario, Haigron,
  Hambly, Harrison, Hernández, Hestroffer, Hodgkin, Janßen, Jevardat~de
  Fombelle, Jordan, Krone-Martins, Lanzafame, Löffler, Marchal, Marrese,
  Moitinho, Muinonen, Osborne, Pancino, Pauwels, Recio-Blanco, Reylé, Riello,
  Rimoldini, Roegiers, Rybizki, Sarro, Smith, Utrilla, van Leeuwen, Abbas,
  Ábrahám, Abreu~Aramburu, Aerts, Aguado, Ajaj, Aldea-Montero, Altavilla,
  Álvarez, Alves, Anders, Anderson, Anglada~Varela, Antoja, Baines, Baker,
  Balaguer-Núñez, Balbinot, Balog, Barache, Barbato, Barros, Bartolomé,
  Bassilana, Becciani, Bellazzini, Berihuete, Bernet, Bertone, Bianchi,
  Binnenfeld, Blanco-Cuaresma, Blazere, Boch, Bombrun, Bossini, Bouquillon,
  Bragaglia, Bramante, Breedt, Bressan, Brouillet, Brugaletta, Bucciarelli,
  Burlacu, Butkevich, Buzzi, Caffau, Cancelliere, Cantat-Gaudin, Carballo,
  Carlucci, Carnerero, Carrasco, Casamiquela, Castellani, Castro-Ginard,
  Chaoul, Charlot, Chemin, Chiaramida, Chiavassa, Chornay, Comoretto, Contursi,
  Cooper, Cornez, Cowell, Crifo, Cropper, Crosta, Crowley, Dafonte, Dapergolas,
  David, de~Laverny, De~Luise, De~March, De~Ridder, de~Souza, de~Torres, del
  Peloso, del Pozo, Delbo, Delgado, Delisle, Demouchy, Dharmawardena, Diakite,
  Diener, Distefano, Dolding, Enke, Fabre, Fabrizio, Fedorets, Fernique,
  Figueras, Fournier, Fouron, Fragkoudi, Gai, Garcia-Gutierrez,
  Garcia-Reinaldos, García-Torres, Garofalo, Gavel, Gerlach, Geyer, Gilmore,
  Girona, Giuffrida, Gomel, Gomez, González-Núñez, González-Santamaría,
  González-Vidal, Granvik, Guillout, Guiraud, Gutiérrez-Sánchez, Guy,
  Hatzidimitriou, Hauser, Haywood, Helmer, Helmi, Sarmiento, Hidalgo,
  Hładczuk, Hobbs, Holland, Huckle, Jardine, Jasniewicz, Jean-Antoine~Piccolo,
  Jiménez-Arranz, Juaristi~Campillo, Julbe, Karbevska, Khanna, Kordopatis,
  Korn, Kóspál, Kostrzewa-Rutkowska, Kruszyńska, Kun, Laizeau, Lambert,
  Lanza, Lasne, Le~Campion, Lebreton, Lebzelter, Leccia, Lecoeur-Taibi, Liao,
  Licata, Lindstrøm, Lister, Livanou, Lobel, Lorca, Loup, Madrero~Pardo,
  Magdaleno~Romeo, Managau, Mann, Manteiga, Marchant, Marconi, Marcos,
  Marcos~Santos, Marín~Pina, Marinoni, Marocco, Marshall, Polo,
  Martín-Fleitas, Marton, Mary, Masip, Massari, Mastrobuono-Battisti,
  McMillan, Messina, Michalik, Millar, Mints, Molina, Molinaro, Molnár,
  Monari, Monguió, Montegriffo, Montero, Mor, Mora, Morbidelli, Morris,
  Muraveva, Murphy, Musella, Nagy, Noval, Ocaña, Ogden, Ordenovic, Osinde,
  Pagani, Pagano, Palaversa, Palicio, Pallas-Quintela, Panahi, Payne-Wardenaar,
  Peñalosa~Esteller, Penttilä, Pichon, Piersimoni, Pineau, Plachy, Plum,
  Poggio, Prša, Pulone, Racero, Ragaini, Rainer, Raiteri, Ramos, Ramos-Lerate,
  Regibo, Richards, Rios~Diaz, Ripepi, Riva, Rix, Rixon, Robichon, Robin,
  Robin, Roelens, Rogues, Rohrbasser, Romero-Gómez, Rowell, Royer, Ruz~Mieres,
  Rybicki, Sáez~Núñez, Sagristà~Sellés, Salguero, Samaras,
  Sanchez~Gimenez, Sanna, Santoveña, Sarasso, Schultheis, Sciacca, Segol,
  Segovia, Semeux, Siddiqui, Siebert, Siltala, Silvelo, Slezak, Slezak, Smart,
  Snaith, Solano, Solitro, Souami, Souchay, Spagna, Spina, Spoto, Steele,
  Steidelmüller, Stephenson, Süveges, Surdej, Szabados, Szegedi-Elek, Taris,
  Taylor, Teixeira, Tolomei, Tonello, Torra, Torra, Torralba~Elipe, Trabucchi,
  Tsounis, Turon, Ulla, Unger, Vaillant, van Dillen, van Reeven, Vanel,
  Vecchiato, Viala, Vicente, Voutsinas, Weiler, Wevers, Wyrzykowski, Yoldas,
  Yvard, Zhao, Zorec, \& Zucker}]{gaia_collaboration_gaia_2022}
{Gaia Collaboration}, Arenou, F., Babusiaux, C., {et~al.} 2022, Gaia {Data}
  {Release} 3: {Stellar} multiplicity, a teaser for the hidden treasure, Tech.
  rep., publication Title: arXiv e-prints ADS Bibcode: 2022arXiv220605595G
  Type: article.
\newblock \url{https://ui.adsabs.harvard.edu/abs/2022arXiv220605595G}

\bibitem[{Gan {et~al.}(2022)Gan, Lin, Wang, Mao, Fouqué, Fan, Bedell, Stassun,
  Giacalone, Fukui, Murgas, Ciardi, Howell, Collins, Shporer, Arnold, Barclay,
  Charbonneau, Christiansen, Crossfield, Dressing, Elliott, Esparza-Borges,
  Evans, Gnilka, Gonzales, Howard, Isogai, Kawauchi, Kurita, Liu, Livingston,
  Matson, Narita, Palle, Parviainen, Rackham, Rodriguez, Rose, Rudat,
  Schlieder, Scott, Vezie, Ricker, Vanderspek, Latham, Seager, Winn, \&
  Jenkins}]{gan_toi-530b_2022}
Gan, T., Lin, Z., Wang, S.~X., {et~al.} 2022, Monthly Notices of the Royal
  Astronomical Society, 511, 83, aDS Bibcode: 2022MNRAS.511...83G.
\newblock \url{https://ui.adsabs.harvard.edu/abs/2022MNRAS.511...83G}

\bibitem[{Ginsburg {et~al.}(2019)Ginsburg, Sipőcz, Brasseur, Cowperthwaite,
  Craig, Deil, Guillochon, Guzman, Liedtke, Lian~Lim, Lockhart, Mommert,
  Morris, Norman, Parikh, Persson, Robitaille, Segovia, Singer, Tollerud,
  de~Val-Borro, Valtchanov, Woillez, {Astroquery Collaboration}, \& {a subset
  of astropy Collaboration}}]{ginsburg_astroquery_2019}
Ginsburg, A., Sipőcz, B.~M., Brasseur, C.~E., {et~al.} 2019, The Astronomical
  Journal, 157, 98.
\newblock \url{http://adsabs.harvard.edu/abs/2019AJ....157...98G}

\bibitem[{{Guillot} {et~al.}(2014){Guillot}, {Ida}, \&
  {Ormel}}]{Guillot_filtering_and_processing_2014}
{Guillot}, T., {Ida}, S., \& {Ormel}, C.~W. 2014, \aap, 572, A72

\bibitem[{Halverson {et~al.}(2016)Halverson, Terrien, Mahadevan, Roy, Bender,
  Stefánsson, Monson, Levi, Hearty, Blake, McElwain, Schwab, Ramsey, Wright,
  Wang, Gong, \& Roberston}]{halverson_comprehensive_2016}
Halverson, S., Terrien, R., Mahadevan, S., {et~al.} 2016, SPIE Proceedings Vol.
  9908, 9908, 99086P.
\newblock \url{http://adsabs.harvard.edu/abs/2016SPIE.9908E..6PH}

\bibitem[{Hartman {et~al.}(2015)Hartman, Bayliss, Brahm, Bakos, Mancini,
  Jordán, Penev, Rabus, Zhou, Butler, Espinoza, de~Val-Borro, Bhatti, Csubry,
  Ciceri, Henning, Schmidt, Arriagada, Shectman, Crane, Thompson, Suc, Csák,
  Tan, Noyes, Lázár, Papp, \& Sári}]{hartman_hats-6b_2015}
Hartman, J.~D., Bayliss, D., Brahm, R., {et~al.} 2015, The Astronomical
  Journal, 149, 166, aDS Bibcode: 2015AJ....149..166H.
\newblock \url{https://ui.adsabs.harvard.edu/abs/2015AJ....149..166H}

\bibitem[{Hatzes(2019)}]{hatzes_iodine_2019}
Hatzes, A.~P. 2019, in The doppler method for the detection of exoplanets,
  2514-3433 (IOP Publishing), 6--1 to 6--16, type: Book chapter.
\newblock \url{https://dx.doi.org/10.1088/2514-3433/ab46a3ch6}

\bibitem[{Henden {et~al.}(2018)Henden, Levine, Terrell, Welch, Munari, \&
  Kloppenborg}]{henden_apass_2018}
Henden, A.~A., Levine, S., Terrell, D., {et~al.} 2018, 232, 223.06, conference
  Name: American Astronomical Society Meeting Abstracts \#232.
\newblock \url{http://adsabs.harvard.edu/abs/2018AAS...23222306H}

\bibitem[{Hildebrand(1983)}]{hildebrand_determination_1983}
Hildebrand, R.~H. 1983, Quarterly Journal of the Royal Astronomical Society,
  24, 267, aDS Bibcode: 1983QJRAS..24..267H.
\newblock \url{https://ui.adsabs.harvard.edu/abs/1983QJRAS..24..267H}

\bibitem[{Hobbs {et~al.}(2022)Hobbs, Shorttle, {Madhusudhan}, \&
  {Nikku}}]{hobbs_molecular_2022}
Hobbs, R., Shorttle, O., {Madhusudhan}, \& {Nikku}. 2022, Monthly Notices of
  the Royal Astronomical Society, doi:10.1093/mnras/stac2106, aDS Bibcode:
  2022MNRAS.tmp.2036H.
\newblock \url{https://ui.adsabs.harvard.edu/abs/2022MNRAS.tmp.2036H}

\bibitem[{Hobson {et~al.}(2023)Hobson, Jordán, Bryant, Brahm, Bayliss,
  Hartman, Bakos, Henning, Almenara, Barkaoui, Benkhaldoun, Bonfils, Bouchy,
  Charbonneau, Cointepas, Collins, Eastman, Ghachoui, Gillon, Goeke, Horne,
  Irwin, Jehin, Jenkins, Latham, Moldovan, Murgas, Pozuelos, Ricker, Schwarz,
  Seager, Srdoc, Striegel, Timmermans, Vanderburg, Vanderspek, \&
  Winn}]{hobson_toi-3235_2023}
Hobson, M.~J., Jordán, A., Bryant, E.~M., {et~al.} 2023, The Astrophysical
  Journal, 946, L4, aDS Bibcode: 2023ApJ...946L...4H.
\newblock \url{https://ui.adsabs.harvard.edu/abs/2023ApJ...946L...4H}

\bibitem[{Hoffman \& Gelman(2014)}]{hoffman_no-u-turn_2014}
Hoffman, M.~D., \& Gelman, A. 2014, Journal of Machine Learning Research, 15,
  1593.
\newblock \url{http://jmlr.org/papers/v15/hoffman14a.html}

\bibitem[{{Howard} \& {Guillot}(2023)}]{Howard_nonidealHHe_2023}
{Howard}, S., \& {Guillot}, T. 2023, \aap, 672, L1

\bibitem[{Howell {et~al.}(2011)Howell, Everett, Sherry, Horch, \&
  Ciardi}]{howell_speckle_2011}
Howell, S.~B., Everett, M.~E., Sherry, W., Horch, E., \& Ciardi, D.~R. 2011,
  The Astronomical Journal, 142, 19, aDS Bibcode: 2011AJ....142...19H.
\newblock \url{https://ui.adsabs.harvard.edu/abs/2011AJ....142...19H}

\bibitem[{Huang {et~al.}(2020)Huang, Vanderburg, Pál, Sha, Yu, Fong,
  Fausnaugh, Shporer, Guerrero, Vanderspek, \& Ricker}]{huang_photometry_2020}
Huang, C.~X., Vanderburg, A., Pál, A., {et~al.} 2020, Research Notes of the
  AAS, 4, 204, publisher: American Astronomical Society.
\newblock \url{https://doi.org/10.3847/2515-5172/abca2e}

\bibitem[{Hunter(2007)}]{hunter_matplotlib_2007}
Hunter, J.~D. 2007, Computing in Science Engineering, 9, 90

\bibitem[{Ida \& Lin(2005)}]{ida_toward_2005}
Ida, S., \& Lin, D. N.~C. 2005, The Astrophysical Journal, 626, 1045.
\newblock \url{http://adsabs.harvard.edu/abs/2005ApJ...626.1045I}

\bibitem[{{Jiang} \& {Ormel}(2023)}]{Jiang_planets_in_rings_2023}
{Jiang}, H., \& {Ormel}, C.~W. 2023, \mnras, 518, 3877

\bibitem[{Jordán {et~al.}(2022)Jordán, Hartman, Bayliss, Bakos, Brahm,
  Bryant, Csubry, Henning, Hobson, Mancini, Penev, Rabus, Suc, de~Val-Borro,
  Wallace, Barkaoui, Ciardi, Collins, Esparza-Borges, Furlan, Gan, Benkhaldoun,
  Ghachoui, Gillon, Howell, Jehin, Fukui, Kawauchi, Livingston, Luque, Matson,
  Matthews, Osborn, Murgas, Narita, Palle, Parvianen, \&
  Waalkes}]{jordan_hats-74ab_2022}
Jordán, A., Hartman, J.~D., Bayliss, D., {et~al.} 2022, The Astronomical
  Journal, 163, 125, aDS Bibcode: 2022AJ....163..125J.
\newblock \url{https://ui.adsabs.harvard.edu/abs/2022AJ....163..125J}

\bibitem[{Kagetani {et~al.}(2023)Kagetani, Narita, Kimura, Hirano, Ikoma,
  Ishikawa, Giacalone, Fukui, Kodama, Gore, Schroeder, Hori, Kawauchi,
  Watanabe, Mori, Zou, Ikuta, Krishnamurthy, Zink, Hardegree-Ullman, Harakawa,
  Kudo, Kotani, Kurokawa, Kusakabe, Kuzuhara, de~Leon, Livingston, Nishikawa,
  Omiya, Palle, Parviainen, Serizawa, Teng, Ueda, \&
  Tamura}]{kagetani_mass_2023}
Kagetani, T., Narita, N., Kimura, T., {et~al.} 2023, Publications of the
  Astronomical Society of Japan, doi:10.1093/pasj/psad031, aDS Bibcode:
  2023PASJ..tmp...48K.
\newblock \url{https://ui.adsabs.harvard.edu/abs/2023PASJ..tmp...48K}

\bibitem[{Kanodia(2023)}]{kanodia_shubham_2023_7685628}
Kanodia, S. 2023, shbhuk/pyastrotools: v0.3, vv0.3,  Zenodo,
  doi:10.5281/zenodo.7685628.
\newblock \url{https://doi.org/10.5281/zenodo.7685628}

\bibitem[{Kanodia \& Wright(2018)}]{kanodia_python_2018}
Kanodia, S., \& Wright, J. 2018, Research Notes of the AAS, 2, 4, publisher:
  American Astronomical Society.
\newblock \url{https://doi.org/10.3847/2515-5172/aaa4b7}

\bibitem[{Kanodia {et~al.}(2023)Kanodia, Mahadevan, Libby-Roberts, Stefansson,
  Cañas, Piette, Boss, Teske, Chambers, Zeimann, Monson, Robertson, Ninan,
  Lin, Bender, Cochran, Diddams, Gupta, Halverson, Hawley, Kobulnicky, Metcalf,
  Parker, Powers, Ramsey, Roy, Schwab, Swaby, Terrien, \&
  Wisniewski}]{kanodia_toi-5205b_2023}
Kanodia, S., Mahadevan, S., Libby-Roberts, J., {et~al.} 2023, The Astronomical
  Journal, 165, 120, aDS Bibcode: 2023AJ....165..120K.
\newblock \url{https://ui.adsabs.harvard.edu/abs/2023AJ....165..120K}

\bibitem[{Kumar {et~al.}(2019)Kumar, Carroll, Hartikainen, \&
  Martin}]{kumar_arviz_2019}
Kumar, R., Carroll, C., Hartikainen, A., \& Martin, O.~A. 2019, The Journal of
  Open Source Software, doi:10.21105/joss.01143.
\newblock \url{http://joss.theoj.org/papers/10.21105/joss.01143}

\bibitem[{Kunimoto {et~al.}(2022)Kunimoto, Daylan, Guerrero, Fong, Bryson,
  Ricker, Fausnaugh, Huang, Sha, Shporer, Vanderburg, Vanderspek, \&
  Yu}]{kunimoto_tess_2022}
Kunimoto, M., Daylan, T., Guerrero, N., {et~al.} 2022, The Astrophysical
  Journal Supplement Series, 259, 33, aDS Bibcode: 2022ApJS..259...33K.
\newblock \url{https://ui.adsabs.harvard.edu/abs/2022ApJS..259...33K}

\bibitem[{{Kurokawa} \& {Inutsuka}(2015)}]{kurokawa_radiusinflation_2015}
{Kurokawa}, H., \& {Inutsuka}, S.-i. 2015, \apj, 815, 78

\bibitem[{Laughlin {et~al.}(2004)Laughlin, Bodenheimer, \&
  Adams}]{laughlin_core_2004}
Laughlin, G., Bodenheimer, P., \& Adams, F.~C. 2004, The Astrophysical Journal
  Letters, 612, L73.
\newblock \url{http://adsabs.harvard.edu/abs/2004ApJ...612L..73L}

\bibitem[{{Lecavelier des Etangs} \& {Lissauer}(2022)}]{Lecavelier_IAUdef_2022}
{Lecavelier des Etangs}, A., \& {Lissauer}, J.~J. 2022, \nar, 94, 101641

\bibitem[{{Lightkurve Collaboration} {et~al.}(2018){Lightkurve Collaboration},
  Cardoso, Hedges, Gully-Santiago, Saunders, Cody, Barclay, Hall, Sagear,
  Turtelboom, Zhang, Tzanidakis, Mighell, Coughlin, Bell, Berta-Thompson,
  Williams, Dotson, \& Barentsen}]{lightkurve_collaboration_lightkurve_2018}
{Lightkurve Collaboration}, Cardoso, J. V. d.~M., Hedges, C., {et~al.} 2018,
  Astrophysics Source Code Library, ascl:1812.013.
\newblock \url{https://ui.adsabs.harvard.edu/abs/2018ascl.soft12013L}

\bibitem[{Liu {et~al.}(2019)Liu, Ormel, \& Johansen}]{liu_growth_2019}
Liu, B., Ormel, C.~W., \& Johansen, A. 2019, Astronomy \&amp; Astrophysics,
  Volume 624, id.A114,
  {\textless}NUMPAGES{\textgreater}14{\textless}/NUMPAGES{\textgreater} pp.,
  624, A114.
\newblock
  \url{https://ui.adsabs.harvard.edu/abs/2019A%26A...624A.114L/abstract}

\bibitem[{Luger {et~al.}(2019)Luger, Agol, Foreman-Mackey, Fleming,
  Lustig-Yaeger, \& Deitrick}]{luger_starry_2019}
Luger, R., Agol, E., Foreman-Mackey, D., {et~al.} 2019, The Astronomical
  Journal, 157, 64.
\newblock \url{https://ui.adsabs.harvard.edu/abs/2019AJ....157...64L}

\bibitem[{Manara {et~al.}(2022)Manara, Ansdell, Rosotti, Hughes, Armitage,
  Lodato, \& Williams}]{manara_demographics_2022}
Manara, C.~F., Ansdell, M., Rosotti, G.~P., {et~al.} 2022, Demographics of
  young stars and their protoplanetary disks: lessons learned on disk evolution
  and its connection to planet formation, Tech. rep., publication Title: arXiv
  e-prints ADS Bibcode: 2022arXiv220309930M Type: article.
\newblock \url{https://ui.adsabs.harvard.edu/abs/2022arXiv220309930M}

\bibitem[{Mandel \& Agol(2002)}]{mandel_analytic_2002}
Mandel, K., \& Agol, E. 2002, The Astrophysical Journal Letters, 580, L171.
\newblock \url{http://adsabs.harvard.edu/abs/2002ApJ...580L.171M}

\bibitem[{{Mankovich} \& {Fuller}(2021)}]{Mankovich_saturn_convection_2021}
{Mankovich}, C.~R., \& {Fuller}, J. 2021, Nature Astronomy, 5, 1103

\bibitem[{Mann {et~al.}(2015)Mann, Feiden, Gaidos, Boyajian, \& von
  Braun}]{mann_how_2015}
Mann, A.~W., Feiden, G.~A., Gaidos, E., Boyajian, T., \& von Braun, K. 2015,
  The Astrophysical Journal, 804, 64.
\newblock \url{http://adsabs.harvard.edu/abs/2015ApJ...804...64M}

\bibitem[{Mann {et~al.}(2019)Mann, Dupuy, Kraus, Gaidos, Ansdell, Ireland,
  Rizzuto, Hung, Dittmann, Factor, Feiden, Martinez, Ruíz-Rodríguez, \&
  Chia~Thao}]{mann_how_2019}
Mann, A.~W., Dupuy, T., Kraus, A.~L., {et~al.} 2019, The Astrophysical Journal,
  871, 63.
\newblock \url{http://adsabs.harvard.edu/abs/2019ApJ...871...63M}

\bibitem[{{McCully} {et~al.}(2018){McCully}, {Volgenau}, {Harbeck}, {Lister},
  {Saunders}, {Turner}, {Siiverd}, \& {Bowman}}]{mccully_2018_BANZAI}
{McCully}, C., {Volgenau}, N.~H., {Harbeck}, D.-R., {et~al.} 2018, in Society
  of Photo-Optical Instrumentation Engineers (SPIE) Conference Series, Vol.
  10707, Software and Cyberinfrastructure for Astronomy V, ed. J.~C. {Guzman}
  \& J.~{Ibsen}, 107070K

\bibitem[{McKinney(2010)}]{mckinney_data_2010}
McKinney, W. 2010, in Proceedings of the 9th {Python} in {Science}
  {Conference}, ed. S.~v.~d. Walt \& J.~Millman, 56 -- 61

\bibitem[{{Michel} {et~al.}(2022){Michel}, {Sadavoy}, {Sheehan}, {Looney}, \&
  {Cox}}]{arnaud_VLA1623_mwvl_2022}
{Michel}, A., {Sadavoy}, S.~I., {Sheehan}, P.~D., {Looney}, L.~W., \& {Cox},
  E.~G. 2022, \apj, 937, 104

\bibitem[{Mollière {et~al.}(2022)Mollière, Molyarova, Bitsch, Henning,
  Schneider, Kreidberg, Eistrup, Burn, Nasedkin, Semenov, Mordasini, Schlecker,
  Schwarz, Lacour, Nowak, \& Schulik}]{molliere_interpreting_2022}
Mollière, P., Molyarova, T., Bitsch, B., {et~al.} 2022, The Astrophysical
  Journal, 934, 74, aDS Bibcode: 2022ApJ...934...74M.
\newblock \url{https://ui.adsabs.harvard.edu/abs/2022ApJ...934...74M}

\bibitem[{Monson {et~al.}(2017)Monson, Beaton, Scowcroft, Freedman, Madore,
  Rich, Seibert, Kollmeier, \& Clementini}]{monson_standard_2017}
Monson, A.~J., Beaton, R.~L., Scowcroft, V., {et~al.} 2017, The Astronomical
  Journal, 153, 96.
\newblock \url{http://adsabs.harvard.edu/abs/2017AJ....153...96M}

\bibitem[{{Montes} {et~al.}(2018){Montes}, {Gonz{\'a}lez-Peinado}, {Tabernero},
  {Caballero}, {Marfil}, {Alonso-Floriano}, {Cort{\'e}s-Contreras},
  {Gonz{\'a}lez Hern{\'a}ndez}, {Klutsch}, \& {Moreno-J{\'o}dar}}]{Montes2018}
{Montes}, D., {Gonz{\'a}lez-Peinado}, R., {Tabernero}, H.~M., {et~al.} 2018,
  \mnras, 479, 1332

\bibitem[{Mordasini {et~al.}(2012)Mordasini, Alibert, Benz, Klahr, \&
  Henning}]{mordasini_extrasolar_2012}
Mordasini, C., Alibert, Y., Benz, W., Klahr, H., \& Henning, T. 2012, Astronomy
  and Astrophysics, 541, A97.
\newblock \url{http://adsabs.harvard.edu/abs/2012A%26A...541A..97M}

\bibitem[{{M{\"u}ller} \& {Helled}(2023)}]{Muller_giantplanet)_review_2023}
{M{\"u}ller}, S., \& {Helled}, R. 2023, \aap, 669, A24

\bibitem[{Müller {et~al.}(2020{\natexlab{a}})Müller, Ben-Yami, \&
  Helled}]{muller_theoretical_2020}
Müller, S., Ben-Yami, M., \& Helled, R. 2020{\natexlab{a}}, The Astrophysical
  Journal, 903, 147, aDS Bibcode: 2020ApJ...903..147M.
\newblock \url{https://ui.adsabs.harvard.edu/abs/2020ApJ...903..147M}

\bibitem[{Müller \& Helled(2021)}]{muller_synthetic_2021}
Müller, S., \& Helled, R. 2021, Monthly Notices of the Royal Astronomical
  Society, 507, 2094, aDS Bibcode: 2021MNRAS.507.2094M.
\newblock \url{https://ui.adsabs.harvard.edu/abs/2021MNRAS.507.2094M}

\bibitem[{Müller {et~al.}(2020{\natexlab{b}})Müller, Helled, \&
  Cumming}]{muller_challenge_2020}
Müller, S., Helled, R., \& Cumming, A. 2020{\natexlab{b}}, Astronomy and
  Astrophysics, 638, A121.
\newblock \url{https://ui.adsabs.harvard.edu/abs/2020A&A...638A.121M/abstract}

\bibitem[{Najita \& Kenyon(2014)}]{najita_mass_2014}
Najita, J.~R., \& Kenyon, S.~J. 2014, Monthly Notices of the Royal Astronomical
  Society, 445, 3315, aDS Bibcode: 2014MNRAS.445.3315N.
\newblock \url{https://ui.adsabs.harvard.edu/abs/2014MNRAS.445.3315N}

\bibitem[{{Ohashi} {et~al.}(2023){Ohashi}, {Tobin}, {J{\o}rgensen}, {Takakuwa},
  {Sheehan}, {Aikawa}, {Li}, {Looney}, {Williams}, {Aso}, {Sharma}, {Sai Insa
  Choi}, {Yamato}, {Lee}, {Tomida}, {Yen}, {Encalada}, {Flores}, {Gavino},
  {Kido}, {Han}, {Lin}, {Narayanan}, {Phuong}, {Santamar{\'\i}a-Miranda},
  {Thieme}, {van't Hoff}, {de Gregorio-Monsalvo}, {Koch}, {Kwon}, {Lai}, {Lee},
  {Plunkett}, {Saigo}, {Hirano}, {Lam}, \& {Mori}}]{Ohashi_eDisk_2023}
{Ohashi}, N., {Tobin}, J.~J., {J{\o}rgensen}, J.~K., {et~al.} 2023, \apj, 951,
  8

\bibitem[{Oliphant(2006)}]{oliphant_numpy_2006}
Oliphant, T. 2006, {NumPy}: {A} guide to {NumPy}, published: USA: Trelgol
  Publishing.
\newblock \url{http://www.numpy.org/}

\bibitem[{Oliphant(2007)}]{oliphant_python_2007}
Oliphant, T.~E. 2007, Computing in Science Engineering, 9, 10

\bibitem[{Pascucci {et~al.}(2016)Pascucci, Testi, Herczeg, Long, Manara,
  Hendler, Mulders, Krijt, Ciesla, Henning, Mohanty, Drabek-Maunder, Apai,
  Szűcs, Sacco, \& Olofsson}]{pascucci_steeper_2016}
Pascucci, I., Testi, L., Herczeg, G.~J., {et~al.} 2016, The Astrophysical
  Journal, 831, 125, aDS Bibcode: 2016ApJ...831..125P.
\newblock \url{https://ui.adsabs.harvard.edu/abs/2016ApJ...831..125P}

\bibitem[{Pollack {et~al.}(1996)Pollack, Hubickyj, Bodenheimer, Lissauer,
  Podolak, \& Greenzweig}]{pollack_formation_1996}
Pollack, J.~B., Hubickyj, O., Bodenheimer, P., {et~al.} 1996, Icarus, 124, 62.
\newblock \url{http://adsabs.harvard.edu/abs/1996Icar..124...62P}

\bibitem[{Pérez \& Granger(2007)}]{perez_ipython_2007}
Pérez, F., \& Granger, B.~E. 2007, Computing in Science and Engineering, 9,
  21.
\newblock \url{https://ipython.org}

\bibitem[{Rabus {et~al.}(2019)Rabus, Lachaume, Jordán, Brahm, Boyajian, von
  Braun, Espinoza, Berger, Le~Bouquin, \& Absil}]{rabus_discontinuity_2019}
Rabus, M., Lachaume, R., Jordán, A., {et~al.} 2019, Monthly Notices of the
  Royal Astronomical Society, 484, 2674, aDS Bibcode: 2019MNRAS.484.2674R.
\newblock \url{https://ui.adsabs.harvard.edu/abs/2019MNRAS.484.2674R}

\bibitem[{Ricker {et~al.}(2014)Ricker, Winn, Vanderspek, Latham, Bakos, Bean,
  Berta-Thompson, Brown, Buchhave, Butler, Butler, Chaplin, Charbonneau,
  Christensen-Dalsgaard, Clampin, Deming, Doty, Lee, Dressing, Dunham, Endl,
  Fressin, Ge, Henning, Holman, Howard, Ida, Jenkins, Jernigan, Johnson,
  Kaltenegger, Kawai, Kjeldsen, Laughlin, Levine, Lin, Lissauer, MacQueen,
  Marcy, McCullough, Morton, Narita, Paegert, Palle, Pepe, Pepper, Quirrenbach,
  Rinehart, Sasselov, Sato, Seager, Sozzetti, Stassun, Sullivan, Szentgyorgyi,
  Torres, Udry, \& Villasenor}]{ricker_transiting_2014}
Ricker, G.~R., Winn, J.~N., Vanderspek, R., {et~al.} 2014, Journal of
  Astronomical Telescopes, Instruments, and Systems, 1, 014003.
\newblock
  \url{https://www.spiedigitallibrary.org/journals/Journal-of-Astronomical-Telescopes-Instruments-and-Systems/volume-1/issue-1/014003/Transiting-Exoplanet-Survey-Satellite/10.1117/1.JATIS.1.1.014003.short}

\bibitem[{Rilinger {et~al.}(2023)Rilinger, Espaillat, Xin, Ribas, Macias, \&
  Luettgen}]{rilinger_determining_2023}
Rilinger, A.~M., Espaillat, C.~C., Xin, Z., {et~al.} 2023, The Astrophysical
  Journal, 944, 66, aDS Bibcode: 2023ApJ...944...66R.
\newblock \url{https://ui.adsabs.harvard.edu/abs/2023ApJ...944...66R}

\bibitem[{Robertson {et~al.}(2019)Robertson, Anderson, Stefansson, Hearty,
  Monson, Mahadevan, Blakeslee, Bender, Ninan, Conran, Levi, Lubar, Cole,
  Dykhouse, Kanodia, Nitroy, Smolsky, Tuggle, Blank, Nelson, Blake, Halverson,
  Henderson, Kaplan, Li, Logsdon, McElwain, Rajagopal, Ramsey, Roy, Schwab,
  Terrien, \& Wright}]{robertson_ultrastable_2019}
Robertson, P., Anderson, T., Stefansson, G., {et~al.} 2019, Journal of
  Astronomical Telescopes, Instruments, and Systems, 5, 015003.
\newblock \url{http://adsabs.harvard.edu/abs/2019JATIS...5a5003R}

\bibitem[{Robitaille {et~al.}(2013)Robitaille, Tollerud, Greenfield,
  Droettboom, Bray, Aldcroft, Davis, Ginsburg, Price-Whelan, Kerzendorf,
  Conley, Crighton, Barbary, Muna, Ferguson, Grollier, Parikh, Nair, Günther,
  Deil, Woillez, Conseil, Kramer, Turner, Singer, Fox, Weaver, Zabalza,
  Edwards, Bostroem, Burke, Casey, Crawford, Dencheva, Ely, Jenness, Labrie,
  Lim, Pierfederici, Pontzen, Ptak, Refsdal, Servillat, \&
  Streicher}]{robitaille_astropy_2013}
Robitaille, T.~P., Tollerud, E.~J., Greenfield, P., {et~al.} 2013, Astronomy \&
  Astrophysics, 558, A33.
\newblock
  \url{https://www.aanda.org/articles/aa/abs/2013/10/aa22068-13/aa22068-13.html}

\bibitem[{Salvatier {et~al.}(2016)Salvatier, Wiecki, \&
  Fonnesbeck}]{salvatier_probabilistic_2016}
Salvatier, J., Wiecki, T.~V., \& Fonnesbeck, C. 2016, PeerJ Computer Science,
  2, e55, publisher: PeerJ Inc.

\bibitem[{Schwab {et~al.}(2016)Schwab, Rakich, Gong, Mahadevan, Halverson, Roy,
  Terrien, Robertson, Hearty, Levi, Monson, Wright, McElwain, Bender, Blake,
  Stürmer, Gurevich, Chakraborty, \& Ramsey}]{schwab_design_2016}
Schwab, C., Rakich, A., Gong, Q., {et~al.} 2016, Proceedings of the SPIE, 9908,
  99087H.
\newblock \url{http://adsabs.harvard.edu/abs/2016SPIE.9908E..7HS}

\bibitem[{Schweitzer {et~al.}(2019)Schweitzer, Passegger, Cifuentes, Béjar,
  Cortés-Contreras, Caballero, del Burgo, Czesla, Kürster, Montes,
  Zapatero~Osorio, Ribas, Reiners, Quirrenbach, Amado, Aceituno,
  Anglada-Escudé, Bauer, Dreizler, Jeffers, Guenther, Henning, Kaminski,
  Lafarga, Marfil, Morales, Schmitt, Seifert, Solano, Tabernero, \&
  Zechmeister}]{schweitzer_carmenes_2019}
Schweitzer, A., Passegger, V.~M., Cifuentes, C., {et~al.} 2019, Astronomy and
  Astrophysics, 625, A68.
\newblock \url{http://adsabs.harvard.edu/abs/2019A%26A...625A..68S}

\bibitem[{Schönrich {et~al.}(2010)Schönrich, Binney, \&
  Dehnen}]{schonrich_local_2010}
Schönrich, R., Binney, J., \& Dehnen, W. 2010, Monthly Notices of the Royal
  Astronomical Society, 403, 1829.
\newblock \url{https://academic.oup.com/mnras/article/403/4/1829/1054839}

\bibitem[{Scott {et~al.}(2018)Scott, Howell, Horch, \&
  Everett}]{scott_nn-explore_2018}
Scott, N.~J., Howell, S.~B., Horch, E.~P., \& Everett, M.~E. 2018, Publications
  of the Astronomical Society of the Pacific, 130, 054502, aDS Bibcode:
  2018PASP..130e4502S.
\newblock \url{https://ui.adsabs.harvard.edu/abs/2018PASP..130e4502S}

\bibitem[{{Selina} {et~al.}(2022){Selina}, {Murphy}, \&
  {Beasley}}]{Selina_ngVLA_2022}
{Selina}, R., {Murphy}, E., \& {Beasley}, A. 2022, in Society of Photo-Optical
  Instrumentation Engineers (SPIE) Conference Series, Vol. 12182, Ground-based
  and Airborne Telescopes IX, ed. H.~K. {Marshall}, J.~{Spyromilio}, \&
  T.~{Usuda}, 121820O

\bibitem[{{Selina} {et~al.}(2018){Selina}, {Murphy}, {McKinnon}, {Beasley},
  {Butler}, {Carilli}, {Clark}, {Erickson}, {Grammer}, {Jackson}, {Kent},
  {Mason}, {Morgan}, {Ojeda}, {Shillue}, {Sturgis}, \&
  {Urbain}}]{selina_ngVLA_2018}
{Selina}, R.~J., {Murphy}, E.~J., {McKinnon}, M., {et~al.} 2018, in Society of
  Photo-Optical Instrumentation Engineers (SPIE) Conference Series, Vol. 10700,
  Ground-based and Airborne Telescopes VII, ed. H.~K. {Marshall} \&
  J.~{Spyromilio}, 107001O

\bibitem[{Sozzetti(2023)}]{sozzetti_dynamical_2023}
Sozzetti, A. 2023, A dynamical mass for {GJ} 463 b: {A} massive super-{Jupiter}
  companion beyond the snow line of a nearby {M} dwarf,  arXiv,
  arXiv:2302.00413 [astro-ph], doi:10.48550/arXiv.2302.00413.
\newblock \url{http://arxiv.org/abs/2302.00413}

\bibitem[{Stassun {et~al.}(2018)Stassun, Oelkers, Pepper, Paegert, De~Lee,
  Torres, Latham, Charpinet, Dressing, Huber, Kane, Lépine, Mann, Muirhead,
  Rojas-Ayala, Silvotti, Fleming, Levine, \& Plavchan}]{stassun_tess_2018}
Stassun, K.~G., Oelkers, R.~J., Pepper, J., {et~al.} 2018, The Astronomical
  Journal, 156, 102.
\newblock \url{http://adsabs.harvard.edu/abs/2018AJ....156..102S}

\bibitem[{Stefànsson {et~al.}(2022)Stefànsson, Mahadevan, Petrovich, Winn,
  Kanodia, Millholland, Maney, Cañas, Wisniewski, Robertson, Ninan, Ford,
  Bender, Blake, Cegla, Cochran, Diddams, Dong, Endl, Fredrick, Halverson,
  Hearty, Hebb, Hirano, Lin, Logsdon, Lubar, McElwain, Metcalf, Monson,
  Rajagopal, Ramsey, Roy, Schwab, Schweiker, Terrien, \&
  Wright}]{stefansson_warm_2022}
Stefànsson, G., Mahadevan, S., Petrovich, C., {et~al.} 2022, The Astrophysical
  Journal, 931, L15, aDS Bibcode: 2022ApJ...931L..15S.
\newblock \url{https://ui.adsabs.harvard.edu/abs/2022ApJ...931L..15S}

\bibitem[{{Tanaka} \&
  {Tsukamoto}(2019)}]{Tanaka_pebble_accretion_classI/0_2019}
{Tanaka}, Y.~A., \& {Tsukamoto}, Y. 2019, \mnras, 484, 1574

\bibitem[{{Tazzari} {et~al.}(2021){Tazzari}, {Testi}, {Natta}, {Williams},
  {Ansdell}, {Carpenter}, {Facchini}, {Guidi}, {Hogherheijde}, {Manara},
  {Miotello}, \& {van der Marel}}]{Tazzari_3mm_Lup_survey_2021}
{Tazzari}, M., {Testi}, L., {Natta}, A., {et~al.} 2021, \mnras, 506, 5117

\bibitem[{{The Theano Development Team} {et~al.}(2016){The Theano Development
  Team}, Al-Rfou, Alain, Almahairi, Angermueller, Bahdanau, Ballas, Bastien,
  Bayer, Belikov, Belopolsky, Bengio, Bergeron, Bergstra, Bisson,
  Bleecher~Snyder, Bouchard, Boulanger-Lewandowski, Bouthillier, de~Brébisson,
  Breuleux, Carrier, Cho, Chorowski, Christiano, Cooijmans, Côté, Côté,
  Courville, Dauphin, Delalleau, Demouth, Desjardins, Dieleman, Dinh, Ducoffe,
  Dumoulin, Ebrahimi~Kahou, Erhan, Fan, Firat, Germain, Glorot, Goodfellow,
  Graham, Gulcehre, Hamel, Harlouchet, Heng, Hidasi, Honari, Jain, Jean, Jia,
  Korobov, Kulkarni, Lamb, Lamblin, Larsen, Laurent, Lee, Lefrancois, Lemieux,
  Léonard, Lin, Livezey, Lorenz, Lowin, Ma, Manzagol, Mastropietro, McGibbon,
  Memisevic, van Merriënboer, Michalski, Mirza, Orlandi, Pal, Pascanu,
  Pezeshki, Raffel, Renshaw, Rocklin, Romero, Roth, Sadowski, Salvatier,
  Savard, Schlüter, Schulman, Schwartz, Vlad~Serban, Serdyuk, Shabanian,
  Simon, Spieckermann, Ramana~Subramanyam, Sygnowski, Tanguay, van Tulder,
  Turian, Urban, Vincent, Visin, de~Vries, Warde-Farley, Webb, Willson, Xu,
  Xue, Yao, Zhang, \& Zhang}]{the_theano_development_team_theano_2016}
{The Theano Development Team}, Al-Rfou, R., Alain, G., {et~al.} 2016, arXiv
  e-prints, arXiv:1605.02688.
\newblock \url{https://ui.adsabs.harvard.edu/abs/2016arXiv160502688T}

\bibitem[{{Tsukamoto} {et~al.}(2017){Tsukamoto}, {Okuzumi}, \&
  {Kataoka}}]{Tsukamoto_apparent_diskmass_2017}
{Tsukamoto}, Y., {Okuzumi}, S., \& {Kataoka}, A. 2017, \apj, 838, 151

\bibitem[{Virtanen {et~al.}(2020)Virtanen, Gommers, Oliphant, Haberland, Reddy,
  Cournapeau, Burovski, Peterson, Weckesser, Bright, van~der Walt, Brett,
  Wilson, Jarrod~Millman, Mayorov, Nelson, Jones, Kern, Larson, Carey, Polat,
  Feng, Moore, Vand~erPlas, Laxalde, Perktold, Cimrman, Henriksen, Quintero,
  Harris, Archibald, Ribeiro, Pedregosa, van Mulbregt, \&
  Contributors}]{virtanen_scipy_2020}
Virtanen, P., Gommers, R., Oliphant, T.~E., {et~al.} 2020, Nature Methods, 17,
  261

\bibitem[{Wright {et~al.}(2010)Wright, Eisenhardt, Mainzer, Ressler, Cutri,
  Jarrett, Kirkpatrick, Padgett, McMillan, Skrutskie, Stanford, Cohen, Walker,
  Mather, Leisawitz, Gautier, McLean, Benford, Lonsdale, Blain, Mendez, Irace,
  Duval, Liu, Royer, Heinrichsen, Howard, Shannon, Kendall, Walsh, Larsen,
  Cardon, Schick, Schwalm, Abid, Fabinsky, Naes, \&
  Tsai}]{wright_wide-field_2010}
Wright, E.~L., Eisenhardt, P. R.~M., Mainzer, A.~K., {et~al.} 2010, The
  Astronomical Journal, 140, 1868.
\newblock \url{http://adsabs.harvard.edu/abs/2010AJ....140.1868W}

\bibitem[{Xin {et~al.}(2023)Xin, Espaillat, Rilinger, Ribas, \&
  Macias}]{xin_measuring_2023}
Xin, Z., Espaillat, C.~C., Rilinger, A.~M., Ribas, A., \& Macias, E. 2023, The
  Astrophysical Journal, 942, 4, aDS Bibcode: 2023ApJ...942....4X.
\newblock \url{https://ui.adsabs.harvard.edu/abs/2023ApJ...942....4X}

\bibitem[{Zechmeister {et~al.}(2018)Zechmeister, Reiners, Amado, Azzaro, Bauer,
  Béjar, Caballero, Guenther, Hagen, Jeffers, Kaminski, Kürster, Launhardt,
  Montes, Morales, Quirrenbach, Reffert, Ribas, Seifert, Tal-Or, \&
  Wolthoff}]{zechmeister_spectrum_2018}
Zechmeister, M., Reiners, A., Amado, P.~J., {et~al.} 2018, Astronomy and
  Astrophysics, 609, A12.
\newblock \url{http://adsabs.harvard.edu/abs/2018A%26A...609A..12Z}

\end{thebibliography}

\end{document}